\documentclass[twoside,english,preprint]{emulateapj}
\usepackage[T1]{fontenc}
\usepackage{amssymb}
\usepackage{graphicx}

\makeatletter

\DeclareRobustCommand{\greektext}{%
  \fontencoding{LGR}\selectfont\def\encodingdefault{LGR}}
\DeclareRobustCommand{\textgreek}[1]{\leavevmode{\greektext #1}}
\DeclareFontEncoding{LGR}{}{}
\DeclareTextSymbol{\~}{LGR}{126}
\providecommand{\tabularnewline}{\\}

\usepackage{apjfonts}
\usepackage[hyperfootnotes=false]{hyperref}
\hypersetup{colorlinks=true,citecolor=blue}
\shorttitle{On the Origins of the Diffuse H$\alpha$ Emission}
\shortauthors{SEON ET AL.}

\makeatother

\usepackage{babel}
\begin{document}

\title{ON THE ORIGINS OF THE DIFFUSE H$\alpha$ EMISSION: IONIZED GAS OR
DUST-SCATTERED H$\alpha$ HALOS?}

\author{Kwang-Il Seon\altaffilmark{1} and Adolf N. Witt\altaffilmark{2}}

\altaffiltext{1}{Korea Astronomy and Space Science Institute, Daejeon 305-348, Korea; email: kiseon@kasi.re.kr}
\altaffiltext{2}{Ritter Astrophysical Research Center, University of Toledo, Toledo, OH 43606, USA} 
\begin{abstract}
It is known that the diffuse H$\alpha$ emission outside of bright
\ion{H}{2} regions not only are very extended, but also can occur
in distinct patches or filaments far from H II regions, and the line
ratios of {[}\ion{S}{2}{]} $\lambda$6716/H$\alpha$ and {[}\ion{N}{2}{]}
$\lambda$6583/H$\alpha$ observed far from bright \ion{H}{2} regions
are generally higher than those in the \ion{H}{2} regions. These
observations have been regarded as evidence against the dust-scattering
origin of the diffuse H$\alpha$ emission (including other optical
lines), and the effect of dust scattering has been neglected in studies
on the diffuse H$\alpha$ emission. In this paper, we reexamine the
arguments against dust scattering and find that the dust-scattering
origin of the diffuse H$\alpha$ emission cannot be ruled out. As
opposed to the previous contention, the expected dust-scattered H$\alpha$
halos surrounding \ion{H}{2} regions are, in fact, in good agreement
with the observed H$\alpha$ morphology. We calculate an extensive
set of photoionization models by varying elemental abundances, ionizing
stellar types, and clumpiness of the interstellar medium (ISM) and
find that the observed line ratios of {[}\ion{S}{2}{]}/H$\alpha$,
{[}\ion{N}{2}{]}/H$\alpha$, and \ion{He}{1} $\lambda$5876/H$\alpha$
in the diffuse ISM accord well with the dust-scattered halos around
\ion{H}{2} regions, which are photoionized by late O- and/or early
B-type stars. We also demonstrate that the H$\alpha$ absorption feature
in the underlying continuum from the dust-scattered starlight (``diffuse
galactic light'') and unresolved stars is able to substantially increase
the {[}\ion{S}{2}{]}/H$\alpha$ and {[}\ion{N}{2}{]}/H$\alpha$
line ratios in the diffuse ISM.
\end{abstract}

\keywords{Galaxy: halo --- \ion{H}{2} regions --- ISM: structure --- radiative
transfer --- scattering}

\section{Introduction}

The diffuse H$\alpha$ emission outside of bright \ion{H}{2} regions
is ubiquitous in late-type galaxies and is generally believed to probe
the warm ionized medium (WIM; also called diffuse ionized gas, DIG),
which may be a major component of the interstellar medium (ISM) of
our Galaxy and other late-type galaxies \citep{Reynolds91,Walterbos96,Dettmar2000,Rand2000,Hidalgo-Gamez2005,Haffner2009}.
It has been argued that the WIM is mainly photoionized by ionizing
radiation (Lyman continuum; Lyc) that leaks out of bright \ion{H}{2}
regions associated with O stars \citep{Mathis1986,Mathis2000,Domgorgen94,Sembach2000,Wood2004,Wood2010},
although it is not clear how the Lyc can penetrate the diffuse \ion{H}{1}
that is observed everywhere in the galaxies \citep[e.g., ][]{Seon2009}.

Previous studies of the diffuse H$\alpha$ and other optical emission
lines concluded that the effect of dust scattering is negligible \citep{Reynolds1985a,Walterbos1994,Ferguson96a,Ferguson96b,Wood1999}.
However, it was recently revealed that the H$\alpha$ excess intensity
in a number of high-latitude clouds is due to scattering by interstellar
dust of H$\alpha$ photons originating elsewhere in the Galaxy \citep{Mattila2007,Lehtinen2010,Witt2010}.
More recently, \citet{Seon2011a,Seon2011b} presented a strong connection
between the diffuse H$\alpha$ and far-ultraviolet (FUV) continuum
backgrounds, which likely suggests a common radiative transfer mechanism
responsible for a substantial fraction of both H$\alpha$ line and
FUV continuum photons. Through a comparison of the H$\alpha$ and
FUV continuum backgrounds, we proposed that the diffuse H$\alpha$
emission at high latitudes may be dominated by late O- and/or early
B-type stars (hereafter, late OB stars) and dust scattering \citep{Seon2011b}.
Furthermore, \citet{Brandt2012} detected emission lines H$\alpha$,
H$\beta$, {[}\ion{N}{2}{]} $\lambda$6583, and {[}\ion{S}{2}{]}
$\lambda$6716 at high Galactic latitudes, whose intensities are correlated
with the 100 $\mu$m intensity, providing an independent confirmation
of the importance of dust-scattered components in the optical emission
lines. If the dust scattering indeed significantly contributes to
the diffuse H$\alpha$ emission and the effect of dust scattering
is not properly taken into account, the energy requirement for the
ionization and volume filling fraction of the WIM will be overestimated.
Also, the widely-applied practice of using the all-sky map of diffuse
H$\alpha$ intensity \citep{Finkbeiner2003} as a template for both
the intensity and small-scale structure of the Galactic foreground
free-free emission in cosmic microwave studies \citep[e.g., ][]{Dickinson2009,Gold2011}
needs to be reevaluated, if dust scattering contributes significantly
to the intensity and spatial structure of diffuse H$\alpha$ light
outside the actual \ion{H}{2} regions.

The recent findings about the importance of dust scattering motivated
us to reexamine the rationales behind previous conclusions against
the dust-scattering origin of the diffuse H$\alpha$ emission. It
was also necessary to confirm our previous results that nebulae photoionized
by late OB stars are capable of reproducing the typical values of
the {[}\ion{N}{2}{]} $\lambda$6583/H$\alpha$ and {[}\ion{S}{2}{]}
$\lambda$6716/H$\alpha$ line ratios.

The rejection of dust scattering as a significant contributor to the
diffuse H$\alpha$ and other optical line emissions is mostly based
on two arguments. The first argument is based on the H$\alpha$ morphology
in the surroundings of \ion{H}{2} regions. \citet{Ferguson96a,Ferguson96b}
assumed that in \emph{R}-band continuum images, bright OB associations
would show \emph{R}-band continuum halos as prominent as the H$\alpha$
halos around \ion{H}{2} regions, if the H$\alpha$ halos around \ion{H}{2}
regions are caused by dust scattering. However, they found no corresponding
halos around bright OB associations in the \emph{R}-band images. Additionally,
light scattered by dust was assumed to be usually concentrated around
\ion{H}{2} regions, but the diffuse H$\alpha$ emission is often
observed in distinct patches or filaments far from \ion{H}{2} regions
\citep[e.g., ][]{Hoopes1996}. Consequently, the extended H$\alpha$
emission outside of bright \ion{H}{2} regions was regarded as evidence
of leakage of Lyc photons rather than as the dust scattering of H$\alpha$
photons. Second, the optical line ratios of {[}\ion{N}{2}{]}/H$\alpha$
and {[}\ion{S}{2}{]}/H$\alpha$ observed outside of \ion{H}{2} regions
have been found to be generally higher than the corresponding ratios
in bright \ion{H}{2} regions \citep{Reynolds1985a,Reynolds1987,Walterbos1994}.
It was thereby concluded that the dust-scattering effect could be
ignored in spite of the fact that bright \ion{H}{2} regions are usually
surrounded by a large amount of molecular gas and dust.

In the present study, by using radiative transfer models, we reexamine
whether these arguments are indeed valid in rejecting the significance
of dust scattering in the diffuse optical emission. First, we argue
that the assertion based on the halo extent or morphology is clearly
wrong. Instead, we show that the global H$\alpha$ morphology in face-on
galaxies can be readily explained by dust scattering. Second, we find
that the observed line ratios of {[}\ion{N}{2}{]}/H$\alpha$ and
{[}\ion{S}{2}{]}/H$\alpha$ outside of \ion{H}{2} regions can be
explained reasonably well by the dust-scattered halos around the \ion{H}{2}
regions photoionized by late OB stars, which is consistent with the
main conclusion of \citet{Seon2011b}. We demonstrate these results
by calculating photoionization models of the radiation-bounded \ion{H}{2}
regions and putting the resulting emissivity distributions of the
optical lines into the radiative transfer models as input sources
for a dust scattering simulation, thereby showing that the dust-scattering
origin of the diffuse H$\alpha$ emission cannot be ruled out. Lastly,
we also point out that the H$\alpha$ absorption line in the dust-scattered
starlight (so called ``diffuse galactic light,'' DGL) and the stellar
continuum of unresolved stars can significantly increase the observed
values of the line ratios.

This paper is organized as follows. In Section 2, we discuss the morphology
and the surface brightness distribution of H$\alpha$ halos around
\ion{H}{2} regions caused by dust scattering. Section 3 investigates
one-dimensional and three-dimensional photoionization models, and
the line ratios resulting from the models. In Section 4, we discuss
the effect of the underlying Balmer absorption line on the optical
line ratios. Additional evidence supporting the results are described
in Section 5. Section 6 summarizes the results.

\section{Dust-Scattered H$\alpha$ Halos}

\subsection{Model}

We performed Monte Carlo radiative transfer simulations of the dust-scattered
H$\alpha$ halos around extended sources. For the dust-scattering
simulation, we used a three-dimensional Monte Carlo radiative transfer
code \citep{Lee2008,Seon2009}, which is similar to those described
in \citet{Wood1999} and \citet{Gordon2001}. The basic algorithm
we use for the radiative transfer is based on the work of \citet{Witt1977}.
The first scattering of the photons is forced to insure that every
photon contributes to the scattered light \citep{Cashwell1959}. The
subsequent scatterings are not forced. The direction into which the
photon is scattered at the \emph{n}th scattering is described by spherical
angles, which are determined from a Henyey-Greenstein phase function.
The scattered light surface brightness image is constructed using
the ``peeling-off'' technique \citep{Yusef-Zadeh1984}. Dust albedo
$a$ and asymmetry factors $g$ for each emission line are adopted
from the Milky Way dust model of \citet{Draine03}.

We assumed a homogeneous dusty medium with $|z|\le100$ pc, $|x|\le500$
pc, and $|y|\le500$ pc to include the whole \ion{H}{2} region. The
source center is at $(x,y,z)=(0,0,0)$. The observer is assumed to
be located at a distance of 3 Mpc, appropriate to NGC\,7793 analyzed
by \citet{Ferguson96b}, from the source center along the $z-$axis.
We considered four dust densities corresponding to the number densities
of hydrogen $n_{{\rm H}}=$ 5, 10, 15, and 20 cm$^{-3}$. The Galactic
gas-to-dust ratio (dust mass per hydrogen nucleon of $1.87\times10^{-26}$
g/H) was assumed for calculation of the dust density. The optical
depth from the source center to the observer is then 0.58, 1.17, 1.76,
and 2.35 at H$\alpha$ for hydrogen density of 5, 10, 15, and 20 cm$^{-3}$,
respectively. The cloud thickness of 100 pc in the $z$-direction
was chosen to have an optical depth at H$\alpha$ of about unity for
the hydrogen density of 10 cm$^{-3}$. The spatial extents of the
dust cloud in the $x$- and $y$-axes were chosen to cover sufficiently
large angular sizes of the dust-scattered halos. Note that the distance
to the source and dust density configuration does not alter the main
results discussed here.

We consider two simple types of H$\alpha$ emissivity profile $\epsilon(x,y,z)$
to obtain general properties of dust-scattered halos, which are independent
of the detailed shape of the emissivity distribution of the photon
source. For the first type, H$\alpha$ photons are uniformly emitted
from a sphere with a radius $r_{{\rm out}}$, e.g., $\epsilon(x,y,z)={\rm constant}$
in $r\le r_{{\rm out}}$ and $\epsilon(x,y,z)=0$ otherwise. In the
second type of model, H$\alpha$ photons are uniformly emitted from
a spherical shell with an outer radius $r_{{\rm out}}$ and a width
of $\Delta r=5$ pc, e.g., $\epsilon(x,y,z)={\rm constant}$ in $r_{{\rm in}}\le r\le r_{{\rm out}}$
and $\epsilon(x,y,z)=0$ otherwise. Here, $r_{{\rm in}}=r_{{\rm out}}-\Delta r$
is the inner radius of the spherical shell. In both types of models,
the outer radius of the source is varied as 10, 20, and 40 pc. The
radial coordinates of H$\alpha$ photons for sphere models are randomly
determined by the formula $r=r_{{\rm out}}\xi^{1/3}$, where $\xi$
is a uniform random variable between 0 and 1, to ensure uniform emissivity.
For the shell models, the radial coordinates of H$\alpha$ photons
are determined by the formula $r=[(r_{{\rm out}}^{3}-r_{{\rm in}}^{3})\xi+r_{{\rm in}}^{3}]^{1/3}$.
Azimuthal and polar coordinates ($\phi$ and $\theta$) of photons
are isotropically determined. Photons are then isotropically emitted
from the locations $(r,\theta,\phi)$.

\subsection{General Properties}

We first describe general properties of the dust-scattered halos.
The surface brightnesses of the H$\alpha$ source should obviously
be proportional to the total luminosity of the source. In the calculations,
we assumed that all sources have the same total luminosity while the
source size and emissivity distribution shape vary. Therefore, the
surface brightness (or intensity) of the source decreases with the
angular size of the source when there is no dust-extinction effect.
In general, the surface brightness would decrease with $\sim\theta_{{\rm prof}}^{-2}$,
where $\theta_{{\rm prof}}$ is the angular size of the source emissivity
profile convolved with the telescope point spread function (including
the effect of binning brightness profile).

\begin{figure*}[t]
\begin{centering}
\includegraphics[scale=0.89]{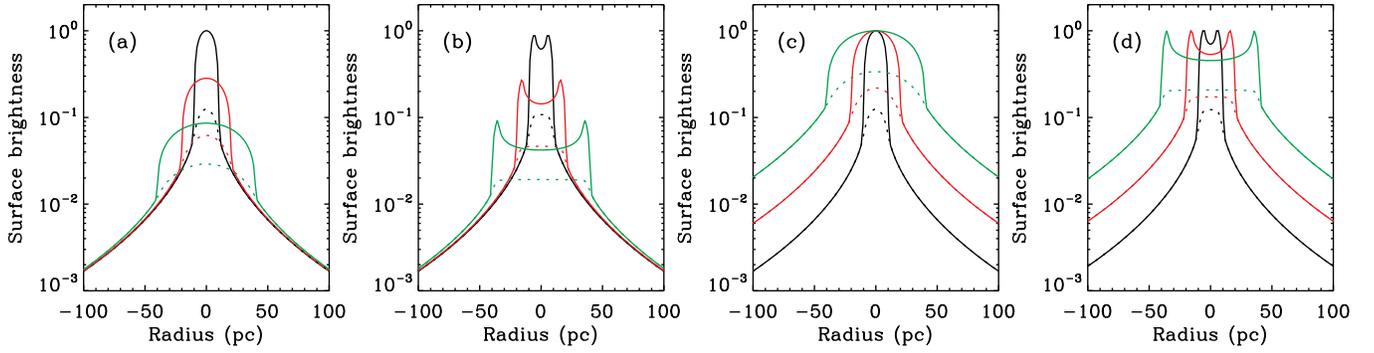}
\par\end{centering}

\caption{\label{ha_halo_size}Dust-scattered halos for various sizes and shapes
of the central source. In the sphere models shown in (a) and (c),
H$\alpha$ photons are uniformly emitted from spheres with radii of
10, 20, and 40 pc. In the shell models shown in (b) and (d), H$\alpha$
photons are uniformly emitted from shells with a radial thickness
of 5 pc and outer radii of 10, 20, and 40 pc. Surface brightnesses
are normalized to the same total luminosity in (a) and (b), and normalized
to the central surface brightness in (c) and (d). Black, red, and
green curves correspond to sources with outer radii of 10, 20, and
40 pc, respectively. Solid and dashed lines denote total (direct +
scattered) and scattered surface brightnesses, respectively.}
\medskip{}
\end{figure*}

When the dust radiative transfer effects are considered, the total
surface brightness (including direct and scattered lights) in the
source region would still decrease as $\theta_{{\rm prof}}$ increases,
although the effects of dust scattering and absorption may alter the
functional dependence of the brightness on $\theta_{{\rm prof}}$.
Figure \ref{ha_halo_size}, in which profiles are binned with a pixel
size of 0.29$''$ (corresponding to 4.2 pc at 3 Mpc), shows the resulting
surface brightness profiles for various source sizes and shapes after
the dust radiative transfer effect was taken into account. In the
figure, hydrogen density of 10 cm$^{-3}$ was assumed. Solid and dashed
lines represent the total (direct + scattered) and scattered surface
brightnesses, respectively. For other hydrogen densities, we obtained
qualitatively similar results as in Figure \ref{ha_halo_size}. The
``surface brightness profile'' $I(r)$ in this paper represents a
horizontal or vertical cut of the two-dimensional surface brightness
distribution across the source center.

Surface brightnesses normalized to the same luminosity are shown in
Figures \ref{ha_halo_size}(a) and (b) for sphere and shell models,
respectively. We note that the surface brightness of the dust-scattered
halo surrounding a larger source declines less steeply near the source
boundary. The most interesting point is that at distances far from
the sources the brightness profiles of the dust-scattered halos for
various sources are essentially the same. This is because an extended
source could be approximated by a point source, regardless of its
shape and size, at locations sufficiently far from the source. Dust
grains at these locations would scatter off the photons as if the
photons had originated from a point source. Therefore, the brightness
profile of the dust-scattered halo at distances far from the source
should be independent of the spatial profile of the source and depend
only on the total luminosity of the source, provided that the dust
configuration (total optical depth and spatial density distribution
of dust) was fixed.

Figures \ref{ha_halo_size}(c) and (d) show surface brightness profiles
normalized to the maximum surface brightness for each source to compare
the relative significance of the dust-scattered halo to the brightness
of \ion{H}{2} region. For a given luminosity, the surface brightness
in the source region decreases as the source size increases, while
the brightness profile of the halo remains relatively invariable.
Therefore, the relative brightness of the scattered halo increases
with the source size, as shown in the figure. We should also note
that the relative contribution of dust-scattered components to the
total intensity at the source region increases with source size. In
Figure \ref{ha_halo_size} ($n_{{\rm H}}=10$ cm$^{-3}$), the contribution
of scattered light to total (direct + scattered) surface brightness
at the center is $\sim$13--46\%. The fraction of the scattered flux
integrated over all regions (\ion{H}{2} + halo regions) to the total
(direct + scattered) flux over all regions is about $\sim$62--63\%.
This ratio is more or less constant, regardless of the source size
and shape, for a given optical depth. The ratio of the scattered flux
integrated over the halo region to the total flux integrated over
all regions is found to be about $\sim$42--57\%. The ratio of scattered
flux in the halo region to the total flux over all regions tends to
decrease slightly with the source size. This is because the scattered
fraction in the \ion{H}{2} region increases with the source size,
while the total scattered-fraction is independent of the size. The
scattered fractions measured in various ways are summarized in Table
\ref{scattered}. Obviously, the contribution of dust-scattered light
to total flux depends on the dust optical depth as well as the geometrical
configuration of the dust cloud. The dust-scattered component becomes
more important when the dust opacity (here, hydrogen density) increases.

\begin{figure}[t]
\begin{centering}
\includegraphics[scale=0.45]{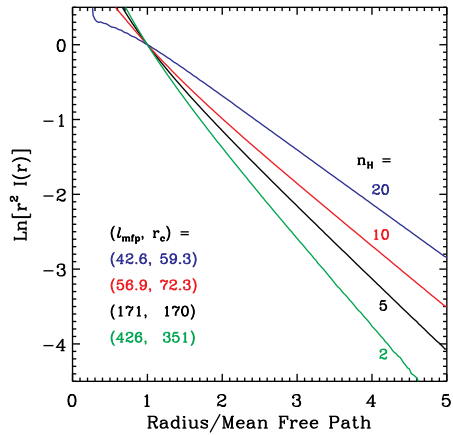}
\par\end{centering}

\caption{\label{ha_halo_profile2}Brightness profiles of dust-scattered halos
for a spherical source with a radius of 10 pc. Hydrogen density varies
from 20 cm$^{-3}$ to 2 cm$^{-3}$ from the upper to lower curves.
The ordinate is shown in natural logarithmic scale to best represent
the slope of exponentially decreasing function. Mean free path $l_{{\rm mfp}}$
and characteristic length $r_{{\rm c}}$, which was obtained by fitting
the brightness profile to a function $r^{-2}\exp(-r/r_{{\rm c}})$,
for each hydrogen density are shown in parentheses.}
\end{figure}

We also note that the surface brightness distribution $I(r)$ of the
individual dust-scattered halos at a large distance $r\gtrsim r_{0}\approx l_{{\rm mfp}}$
can be represented well by the following formula: 
\begin{equation}
I(r)=I(r_{0})\left(\frac{r_{0}}{r}\right)^{2}e^{-(r-r_{0})/r_{{\rm c}}}.\label{eq:halo_profile}
\end{equation}
Here, $I(r_{0})$ and $l_{{\rm mfp}}=(n_{{\rm H}}\sigma_{{\rm ext}})^{-1}$
are the surface brightness at $r=r_{0}$ and the mean free path, respectively.
$\sigma_{{\rm ext}}$ is the dust extinction cross-section per hydrogen.
Figure \ref{ha_halo_profile2} shows the $r^{2}I(r)$ profiles of
the dust-scattered halos surrounding a spherical source with a radius
of 10 pc, in linear-logarithmic scale, in media with various dust
densities corresponding to $n_{{\rm H}}=$ 2, 5, 10, and 20 cm$^{-3}$,
which are typical in the ISM. We fitted the asymptotic profiles at
$r>l_{{\rm mfp}}$, and obtained the best-fit values of the characteristic
scale length $r_{{\rm c}}$, which are consistent with the mean free
path within a factor of 1.4. The mean free paths and the best-fit
scale lengths are shown in the figure. Higher density resulted in
a more extended profile than that expected by $l_{{\rm mfp}}$ due
to multiple scattering. The formula will be used to understand the
increase of the {[}\ion{N}{2}{]}/H$\alpha$ and {[}\ion{S}{2}{]}/H$\alpha$
line ratios with decreasing H$\alpha$ intensity in Section 4 as well
as the global H$\alpha$ morphology in face-on galaxies in Section
5.

Note that our results were inferred from two different spatial distributions
of the photon source, i.e., sphere and shell, and therefore they are
independent of the spatial distribution of H$\alpha$ emission in
\ion{H}{2} regions. We confirmed the results again using more realistic
emissivity distributions predicted from photoionization models in
Section 3. We also note that scattering by dust within \ion{H}{2}
regions is not a major concern of the present study. The halos, which
we are mainly interested in, are the results of scattering by dust
in the lines of sight passing through the ISM outside of \ion{H}{2}
regions. Dust within \ion{H}{2} regions contributes to the scattered
light that is observed in the sightlines toward \ion{H}{2} regions.

\subsection{Comparison with Observations}

\citet{Ferguson96a} found that there is no evidence for halos around
bright OB associations in the \emph{R}-band continuum images, which
they assume to be equally extended as the H$\alpha$ halos surrounding
\ion{H}{2} regions if the H$\alpha$ halos are indeed the result
of dust scattering. Therefore, they concluded that the role of the
dust scattering is negligible in the H$\alpha$ emission surrounding
bright \ion{H}{2} regions. We here note that the H$\alpha$ sources
are mainly concentrated in relatively few bright \ion{H}{2} regions
of substantial H$\alpha$ luminosity. On the other hand, the diffuse
\emph{R}-band continuum background in galaxy images is far more uniformly
distributed spatially, because the \emph{R}-band continuum sources
do not just consist of OB associations but are made up mostly by later-type
stars, including late-type giants, which have a much more uniform
distribution than that of the OB associations. Accordingly, the dust-scattered
halos surrounding OB associations in the \emph{R}-band images would
be hidden among the diffuse background, composed of widely distributed
stellar sources and the dust-scattered light produced by them.

\begin{figure}[t]
\begin{centering}
\includegraphics[scale=0.89]{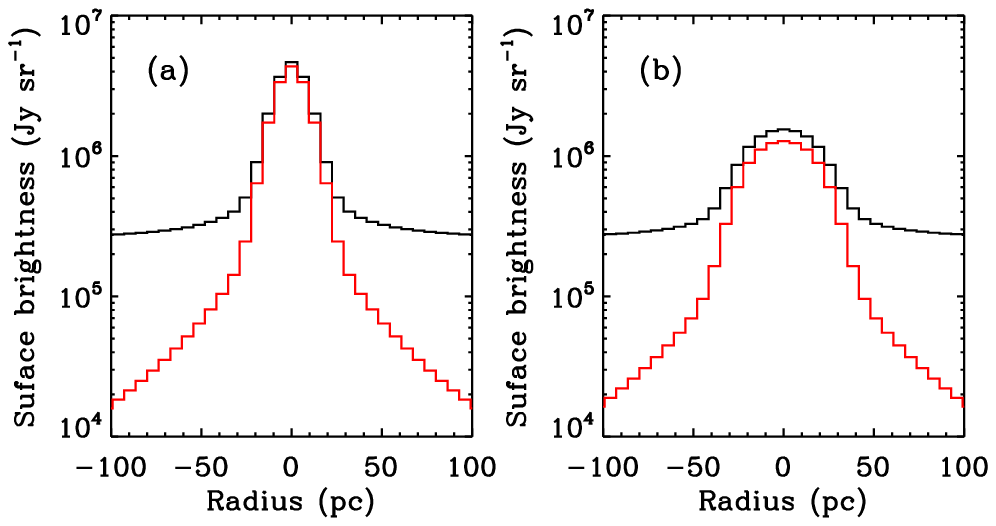}\includegraphics[scale=0.89]{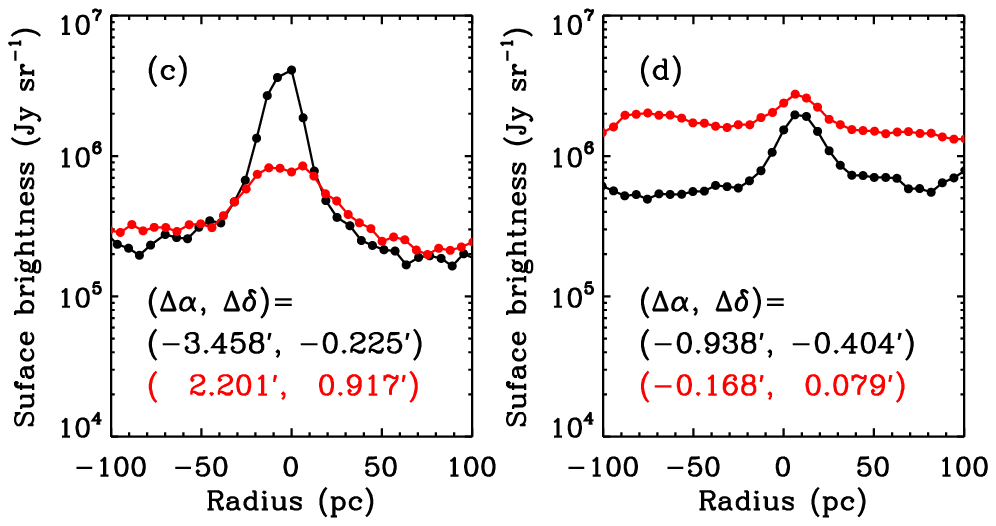}
\par\end{centering}

\caption{\label{Rband}Surface brightness profiles of OB associations in \emph{R}-band
image. Model profiles for OB association with radii of (a) 10 pc and
(b) 30 pc, which are expected from a telescope with a spatial resolution
of 1.2$''$. Black curves simulate brightness profiles of OB associations
in the presence of diffuse background. The expected profiles in the
case of no background are shown in red color. Hydrogen density was
10 cm$^{-3}$ for the models. Four sample profiles of OB associations
in NGC\,7793, taken from SINGS data, are shown in (c) and (d). In
the figures (c) and (d), the coordinates $(\Delta\alpha,\Delta\delta)$
of two OB associations relative to the galactic center $(\alpha,\delta)$
= (23$^{{\rm h}}$57$^{{\rm m}}$49.83$^{{\rm s}}$, $-$32$^{{\rm d}}$35$^{{\rm m}}$27.7$^{{\rm s}}$)
are shown in units of arcmin in parentheses.}
\end{figure}

Typical H$\alpha$ luminosities of bright \ion{H}{2} regions, for
example, in NGC\,7793 analyzed by \citet{Ferguson96b} are $\gtrsim10^{39}$
erg s$^{-1}$. Assuming a typical H$\alpha$ luminosity of $2\times10^{39}$
erg s$^{-1}$ and case B recombination, Lyc luminosity of a typical
OB association is $Q({\rm H}^{0})=1.4\times10^{51}$ photons s$^{-1}$,
which is equivalent to the luminosity produced by 160 stars of O7V
type (see Table \ref{sp_type} which will be explained in Section
3). Using the Kurucz stellar model \citep{Castelli03}, \emph{R}-band
luminosity of the OB association composed of 160 O7V stars is $7\times10^{24}$
erg s$^{-1}$ Hz$^{-1}$. We verified the estimated luminosities with
an evolutionary synthesis code Starburst99 \citep{Leitherer1999,Leitherer2010}.
Assuming a Salpeter initial mass function with a total mass of $10^{4}M_{\odot}$
in the mass range of $6M_{\odot}\le M\le100M_{\odot}$, an 1.3 Myr-old
OB association produces the same Lyc and \emph{R}-band luminosities
as the above Kurucz model. Adopting the total \emph{R}-band magnitude
of 8.71 mag, corresponding to 1.01 Jy (1 Jy = $10^{-23}$ erg s$^{-1}$
cm$^{-2}$ Hz$^{-1}$), and the major and minor diameters of 9.3$'$
and 6.3$'$, taken from NASA/IPAC Extragalactic Database (NED), the
average background surface brightness in \emph{R}-band is $2.6\times10^{5}$
Jy sr$^{-1}$. The surface brightness of the galaxy shows a roughly
exponential decrease with a scalelength of $1'.1$ ($\sim1$ kpc)
\citep{Carignan1985}. A similar brightness to the above average value
is found at the galactocentric radius of $\sim3'$ from the radial
profile. Using these values, we calculated the expected surface brightness
profiles in \emph{R}-band for OB associations with radii of 10 and
30 pc, as shown in Figures \ref{Rband}(a) and (b), respectively.
In the calculations, we assumed that photon sources are uniformly
distributed within the assumed radii and the hydrogen density was
10 cm$^{-3}$. The calculated images were convolved with a two-dimensional
gaussian function with a full width at half maximum of $1.2''$ (17
pc at 3 Mpc) and the brightness profiles were obtained with a bin
size of $0.44''$ to match the seeing and pixel size of the observations
of \citet{Ferguson96b}. Black curves show the resulting surface brightness
profiles in the presence of the diffuse background, while red curves
show the cases where there is no constant background. It is now clear
that the dust-scattered halos of OB associations in \emph{R}-band
images are not observable. Red curves indicate that the dust-scattered
halos can be observed only when there is no diffuse background, as
in the case of H$\alpha$ emission originating from relatively few
\ion{H}{2} regions.

We also examined the \emph{R}-band image of NGC\,7793 observed by
the SIRTF Nearby Galaxy Survey (SINGS) \citep{Kennicutt2003,Dale2007},
which were obtained with the same telescope as used by \citet{Ferguson96b}.
Figures \ref{Rband}(c) and (d) show examples of the observed \emph{R}-band
brightness profile around OB associations located at the coordinates
$(\Delta\alpha,\Delta\delta)$, denoted in the figures, relative to
the galactic center. In Figure \ref{Rband}(c), of which the samples
are located at a galactocentric radius of $\sim3'$, the black curve
represents a typical sample of OB associations with the highest brightness
contrast compared to their surrounding background, while the red curve
shows a case with relatively low contrast. Figure \ref{Rband}(d)
shows two more OB associations that are located closer to the galactic
center. Because of a higher background level, the profiles near the
galactic center have lower contrast. In general, brightness profiles
of OB associations are more or less similar to one of the four examples
in the figures, after multiplying a scale factor depending on location.
The profiles tend to have sharper contrast at larger galactic radius,
but not always. We note that the models in Figures \ref{Rband}(a)
and (b) reproduce the observed profiles of OB associations in Figure
\ref{Rband}(c) perfectly. Our models are also able to reproduce the
brightness profiles shown in Figure \ref{Rband}(d) by varying the
background level and the size of OB associations We, therefore, conclude
that the presence of a continuum background washes out any observable
trace of the scattered continuum halo of OB associations.

The next fact we need to note is that dust clouds far from \ion{H}{2}
regions can scatter H$\alpha$ photons originating elsewhere in galaxies,
as demonstrated for high-latitude clouds in the Milky Way \citep{Mattila2007,Lehtinen2010,Witt2010}.
Filaments, patches, and arcs, which are often observed not only in
the Milky Way but also in external galaxies, do not necessarily indicate
in situ ionized gas. Unlike the assertion of \citet{Hoopes1996},
these distinct features can be explained by dust scattering as well.
In the halos of edge-on galaxies, very little correspondence between
the H$\alpha$-emitting filaments and ``absorbing'' dust structures
was found \citep{HowkSavage2000}, which seems to indicate the physically
distinct origins of dust and H$\alpha$-emitting material. However,
we should note that the scattered intensity toward a dust cloud can
be higher, lower, or equal to its surroundings (as demonstrated in
\citealp{Mattila2007}), depending on the incident intensity from
behind of the cloud and the cloud optical depth. A cloud with an optical
depth $\gtrsim1-2$ can appear dark, while a cloud with a lower optical
depth appears bright. Indeed, no correlation and anti-correlation
between the dust-scattered FUV continuum background and dust has been
found at high opacity regions \citep[e.g., ][]{Seon2011a}. As noted
by \citet{HowkSavage2000}, the dust filamentary features observed
in edge-on galaxies are traced in absorption against the background
starlight, thereby implying relatively opaque dust clouds $(A_{{\rm V}}\gtrsim0.8-2.0)$.
Therefore, no correspondence between the H\textgreek{a}-emitting filaments
and absorbing dust features does not necessarily indicate the distinct
origins of the dust and H$\alpha$ filaments, but can also be a natural
consequence of the combined effect of absorption and scattering at
relatively high opacity.

It is also well known from the observations of face-on galaxies that
the radial and azimuthal intensity distributions of the diffuse H$\alpha$
emission are highly correlated with bright \ion{H}{2} regions over
both small and large scales (e.g., \citealt{Walterbos1994,Ferguson96a,Ferguson96b,Zurita02}).
In fact, the strong correlation of the halo brightness with that of
the \ion{H}{2} region is a natural consequence of dust scattering.
The mean fraction of the diffuse H$\alpha$ emission outside of \ion{H}{2}
regions is known to be roughly constant ($\approx0.5$), regardless
of Hubble type and the star formation rate per unit disk area \citep{Ferguson96a,Ferguson96b,Hoopes1996,Greenawalt1998,Zurita00,Voges2006,Oey2007}.
This mean value matches well with the dust-scattered fraction estimated
in the present study, as shown in Table \ref{scattered}. In fact,
it is easy to construct cases where the reflected nebular flux surrounding
a star is comparable or even larger than the source flux (see Fig.~9
in \citealp{Witt1982}).

\citet{Zurita00} found a weak tendency for the diffuse H$\alpha$
fraction to increase with the galactocentric radius. The weak trend
can be understood qualitatively, if the diffuse H$\alpha$ emission
is dominated by dust scattering. Sizes of \ion{H}{2} regions would
generally decrease with the galactocentric radius \citep[e.g., ][]{FichBlitz1984}.
As we noted previously (Table \ref{scattered}), the dust-scattered
fraction in the halo region slightly increases as the source size
decreases. Therefore, the scattered diffuse H$\alpha$ fraction may
slightly increase with the galactocentric radius.

Here, we considered H$\alpha$ halos around single \ion{H}{2} regions.
In a realistic case, the dust-scattered halos from adjacent \ion{H}{2}
regions would merge into each other and produce the pervasive H$\alpha$
background surrounding \ion{H}{2} regions. Therefore, we may need
more detailed radiative transfer models for individual face-on galaxies
as a whole system to investigate which mechanism, the dust-scattering
of H$\alpha$ or in situ photoionization by Lyc leakage, better explains
the overall H$\alpha$ morphologies of face-on galaxies. It is worthwhile
to note that some quantitative analyses on the H$\alpha$ morphology
were already preformed by \citet{Zurita02} and \citet{Seon2009}.
In Section 5, we will show that their analyses for the H$\alpha$
morphologies of face-on galaxies likely better support the dust-scattered
H$\alpha$ scenario than the Lyc leakage scenario.

In summary, the previous arguments based on the spatial extent or
morphology of H$\alpha$ halos to rule out the dust-scattering origin
of the diffuse H$\alpha$ emission are no longer valid. Instead, dust
scattering explains the general properties of the diffuse H$\alpha$
morphology very well.

\section{Optical Line ratios}

\subsection{Basic Assumptions}

We used the well-known one-dimensional photoionization code CLOUDY
(version c10.00 of the code last described by \citealt{Ferland1998})
and the fully three-dimensional Monte Carlo photoionization code MOCASSIN
(MOnte CArlo SimulationS of Ionized Nebulae; \citealt{Ercolano2003,Ercolano2005,Ercolano2008})
to model spherically-symmetric radiation-bounded \ion{H}{2} regions
surrounding single ionizing stars in a homogeneous medium. We also
modeled ionized nebulae in a clumpy medium by using the MOCASSIN code.

Extensive calculations of \ion{H}{2} region models were performed
by varying elemental abundances, spectral type of the central ionizing
source, and hydrogen number density. Various sets of elemental abundances
have been previously used in modeling the WIM. In the present study,
we used six sets of abundances, which are summarized in Table \ref{abundance}:
new solar system, B star, Orion nebula, ISM, warm neutral medium (WNM)
values, and WNM values with enhanced S abundance (hereafter, WNM2).
The abundance of He is assumed to be 0.1 of the abundance of H in
all six cases. The new solar system, ISM, and Orion nebula abundances
are as defined in CLOUDY. The abundances for B star and WNM are defined
in \citet{Sembach2000}. The WNM abundances represent those observed
in the well-studied warm diffuse clouds toward the low-halo star $\mu$
Columbae. The WNM abundances were also used by \citet{Mathis2000}.
The WNM2 abundances were used by \citet{Wood2004} and are essentially
the same as the WNM abundances except that the S abundance is increased
by 0.2 dex. We note that the new solar abundances adopted in this
paper are different from the old solar abundances used in \citet{Sembach2000}.
For the model calculations using CLOUDY, we assumed Orion dust grains
for Orion abundances and ISM dust grains for the others, as defined
in CLOUDY. In the MOCASSIN models, no grains were assumed. The effect
of dust grains is mainly to reduce the ionized region. However, no
significant effect on the line ratios is found \citep{Mathis1986b}.
We ignored elements other than those shown in Table \ref{abundance}.
The electron temperature and ionization structure of \ion{H}{2} regions
are mainly determined by H, He, N, O, Ne, and S, which are major coolants
\citep[e.g., ][]{Mathis1985}. The elements that were not included
in the calculations have no significant effect on the models of \ion{H}{2}
regions.

Spectra of ionizing stars are also needed to model photoionized nebulae.
We considered only dwarf stars with spectral types from O3V to B1V.
For each spectral type, we adopted a stellar temperature, radius,
and effective gravity derived from the evolutionary tracks in \citet{Straizys81}.
We calculated the spectral energy distribution and the hydrogen ionization
luminosity $Q({\rm H}^{0})$ for each spectral type with a grid of
Kurucz models, which are plane-parallel line-blanketed model atmospheres
in local thermodynamic equilibrium (LTE) \citep{Castelli03}. Table
\ref{sp_type} summarizes the stellar temperatures, radii $R_{*}$,
H-ionizing (Lyc) luminosities $Q({\rm H}^{0})$, He-ionizing luminosities
$Q({\rm He}^{0})$, and gravities $g_{*}$ for the spectral types
used in this study. The non-LTE (NLTE) WMBASIC model atmospheres \citep{Pauldrach2001,Sternberg2003}
are also used to calculate the one-dimensional \ion{H}{2} regions
and the results are compared with those obtained with the LTE atmospheres.
Most of the photoionization models in the present study are calculated
by using Kurucz's LTE atmospheres, unless otherwise specified.

The model calculations using CLOUDY were terminated when the fraction
of neutral hydrogen ($x_{{\rm edge}}$) reached 0.99, at which point
the emissivities are very low. We found no significant difference
when further calculations were performed beyond this value. Note that
most of the one-dimensional WIM models used nebulae models with $x_{{\rm edge}}=0.95$
\citep{Mathis1986} or combinations of models with $x_{{\rm edge}}=0.10$
and $x_{{\rm edge}}=0.95$ to explain line ratios \citep{Domgorgen94,Sembach2000,Mathis2000}.
For the models calculated with MOCASSIN, the sizes of density grid
structures were decided to be at least $\sim$20\% larger than the
Str{\"o}mgren radius calculated with CLOUDY to take the effect of
absence of dust into account and ensure the models cover the whole
ionized regions. Reflection symmetry in the $x$-, $y$-, and $z$-directions
was assumed for fast calculations. The number of grids is $33\times33\times33$
for the MOCASSIN models.

We present only the emission lines H$\alpha$, {[}\ion{N}{2}{]} $\lambda$6583,
and {[}\ion{S}{2}{]} $\lambda$6716, which were best investigated
in the Milky Way, and \ion{He}{1} $\lambda$5876, which can best
constrain the spectral type of ionizing stars. The volume emissivities
of H$\alpha$, \ion{He}{1}, {[}\ion{N}{2}{]}, and {[}\ion{S}{2}{]}
lines obtained using the CLOUDY and MOCASSIN models were used as input
emissivity distributions to calculate the dust-scattered halos of
the optical lines surrounding \ion{H}{2} regions. For the dust halo
models of spherically-symmetric \ion{H}{2} regions calculated with
the CLOUDY code, the initial radial coordinates of photons were determined
by the volume emissivity $\epsilon$ of each optical line at a radius
$r$, i.e., $P(r)dr=\epsilon r^{2}dr/\int\epsilon r^{2}dr$. In the
dust-scattering models of the \ion{H}{2} regions calculated with
the MOCASSIN code, the initial coordinates $(x,y,z)$ of photons were
randomly selected to follow the volume emissivity $\epsilon(x,y,z)$,
i.e., $P(x,y,z)dV=\epsilon(x,y,z)dV/\int\epsilon(x,y,z)dV$, and their
initial directions were randomly chosen from an isotropic distribution.

\subsection{One-dimensional Models}

\begin{figure*}[t]
\begin{centering}
\includegraphics[scale=0.89]{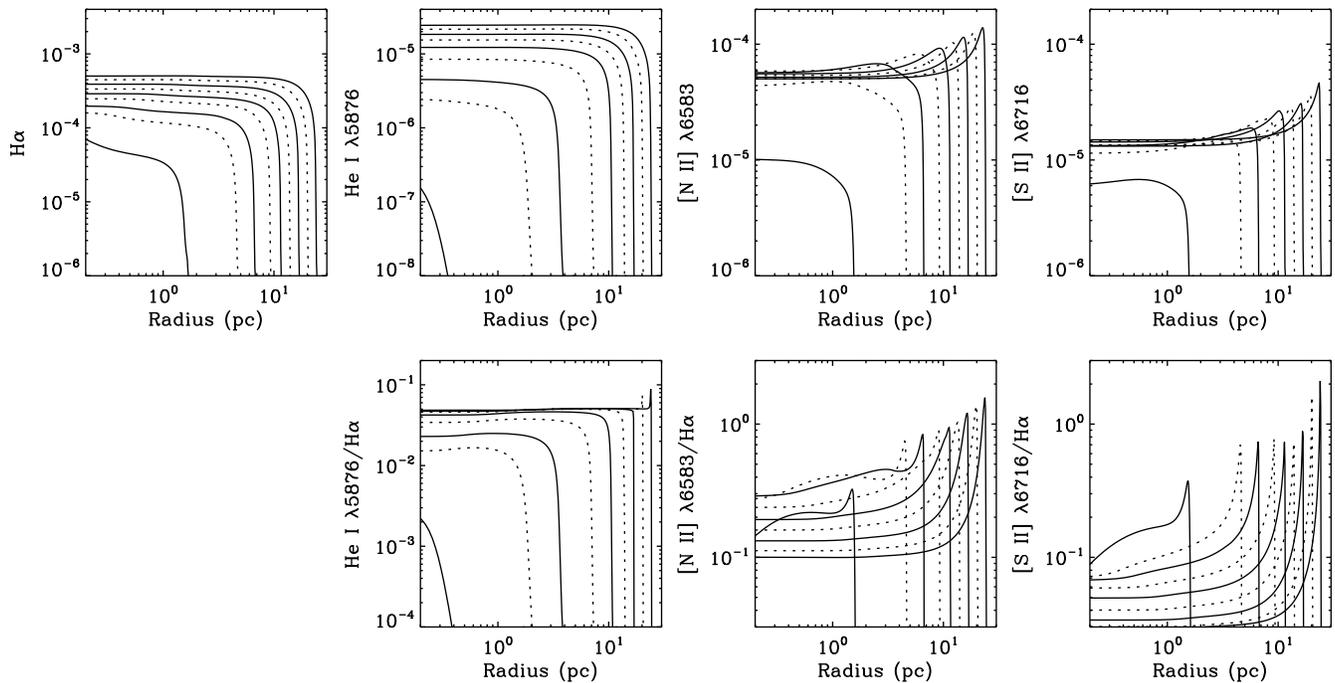}
\par\end{centering}

\caption{\label{intensity_profile}Top: brightness profiles of H$\alpha$,
\ion{He}{1} $\lambda$5867, {[}\ion{N}{2}{]} $\lambda6583$, and
{[}\ion{S}{2}{]} $\lambda$6716 lines (in units of erg cm$^{-2}$
s$^{-1}$ sr$^{-1}$) for various central ionization sources. Bottom:
brightness profiles of line ratios \ion{He}{1}/H$\alpha$, {[}\ion{N}{2}{]}/H$\alpha$,
and {[}\ion{S}{2}{]}/H$\alpha$. Elemental abundances for WNM and
hydrogen density of $n_{{\rm H}}=10$ cm$^{-3}$ were assumed for
the photoionization models. The curves from the outermost to innermost
correspond to O3V to B1V stars progressively. Solid and dashed lines
were alternatively used for clarification.}
\end{figure*}

We first modeled the one-dimensional \ion{H}{2} regions by varying
hydrogen number density $n_{{\rm H}}$, elemental abundances, and
spectral type of the central ionizing source. Hydrogen number density
was varied as 0.1, 1, 10, and 100 cm$^{-3}$. Figure \ref{intensity_profile}
shows brightness profiles of line intensities (H$\alpha$, \ion{He}{1},
{[}\ion{N}{2}{]}, and {[}\ion{S}{2}{]}) and intensity ratios (\ion{He}{1}/H$\alpha$,
{[}\ion{N}{2}{]}/H$\alpha$, and {[}\ion{S}{2}{]}/H$\alpha$), projected
onto the sky, for various spectral types of ionizing source. In the
figure, the hydrogen density and elemental abundances were assumed
to be 10 cm$^{-3}$ and WNM, respectively. The curves from the outermost
to innermost correspond to later type stars progressively. The outermost
curve, corresponding to the largest \ion{H}{2} region, is obtained
from an O3V star, and the innermost curve corresponds to a B1V star.

Emissivities of the low ionization lines {[}\ion{S}{2}{]} and {[}\ion{N}{2}{]}
increase suddenly near the ionization boundary of the \ion{H}{2}
region, for most of ionizing stellar types except B1V, and then drop,
whereas the H$\alpha$ and \ion{He}{1} recombination lines (RLs)
decrease continuously. In other words, both {[}\ion{N}{2}{]} and
{[}\ion{S}{2}{]} arise primarily in the transition region between
the ionized and partially ionized zones of the \ion{H}{2} region
\citep[e.g., ][]{Evans1985}. Therefore, the line ratios of {[}\ion{S}{2}{]}/H$\alpha$
and {[}\ion{N}{2}{]}/H$\alpha$ increase rapidly at the ionization
boundary. Such a rapid increase in the line ratios at the ionization
boundary is shown in the Orion nebula \citep{Pogge1992,Sanchez2007,Mesa-Delgado2011}.
The giant \ion{H}{2} region NGC\,595 in M\,33 also shows an increase
in the line ratios at the ionization boundary \citep{Relano2010}.

The line ratio {[}\ion{N}{2}{]}/H$\alpha$ shows a less rapid increase
at the boundaries than {[}\ion{S}{2}{]}/H$\alpha$. S$^{+}$ is much
more confined to the outer regions of the Str{\"o}mgren sphere than
is N$^{+}$ \citep{Mathis2000}. The reason is that the change in
the predominant ionization state between the central and outer \ion{H}{2}
regions is smaller for N because of its higher ionization potential
for N$^{+}$ $\rightarrow$ N$^{2+}$: the ionization potentials for
N$^{+}$ $\rightarrow$ N$^{2+}$ and S$^{+}$ $\rightarrow$ S$^{2+}$
are 29.6 eV and 23.3 eV, respectively, and therefore S exists in the
doubly ionized state out to larger radius than N.

For the hottest (O3V and O4V) stars, which emit a sufficient number
of photons at higher energies than the He$^{0}$ ionization potential
(24.6 eV), photons with $E$ < 24.6 eV run out slightly faster than
those with $E$ > 24.6 eV at the ionization boundary. This resulted
in a less rapid decline of the \ion{He}{1} line intensity than H$\alpha$
and a weak spike in the line ratio \ion{He}{1}/H$\alpha$ at the
ionization boundary, as shown in Figure \ref{intensity_profile}.
However, for later stars, He becomes neutral at the inner zone relative
to H, and thereby \ion{He}{1}/H$\alpha$ line declines rapidly at
a much smaller radius than the Str{\"o}mgren radius.

As already noted, the surface brightness of the dust-scattered halo
depends on the total luminosity of the source only, for a fixed dust
configuration. Therefore, the intensity ratio of a pair of emission
lines in the dust-scattered halo is expected to be determined by the
ratio of total luminosities of the lines, rather than by the ratio
of surface brightnesses at the main part of the \ion{H}{2} region,
if their wavelengths are close enough to each other. If their wavelengths
are different to the extent that dust extinction cross sections and
albedos are significantly different, their ratio in the dust-scattered
halo would also differ from the luminosity ratio. In the brightness
profiles, the highest {[}\ion{N}{2}{]} and {[}\ion{S}{2}{]} intensities
occur in relatively small regions at the ionization boundary. However,
the transition zones can contribute significantly to total luminosities
because the luminosities are defined by integrals $4\pi\int\epsilon r^{2}dr$
and the highest emissivities of {[}\ion{N}{2}{]} and {[}\ion{S}{2}{]}
are found at the largest radii. Therefore, the line ratios of {[}\ion{S}{2}{]}/H$\alpha$
and {[}\ion{N}{2}{]}/H$\alpha$ at the dust-scattered halo should
be larger than the values at the central part of the \ion{H}{2} region,
but lower than the values at the boundary as a result of dust-scattering.

\begin{figure*}[t]
\begin{centering}
\includegraphics[scale=0.89]{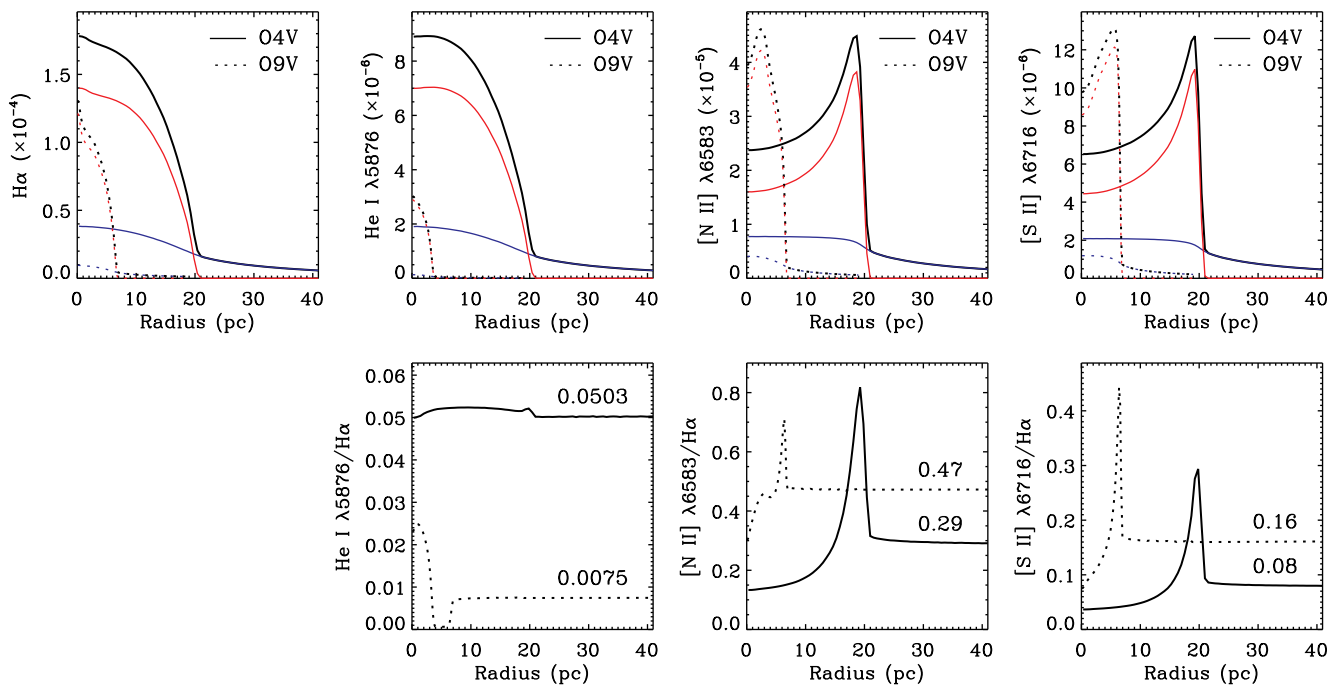}
\par\end{centering}

\caption{\label{scatt_ratio_O4}Brightness profiles of dust-scattered optical
emission lines originating from the \ion{H}{2} region surrounding
O4V (solid lines) and O9V (dotted lines) stars. Top panels from left
to right show intensity profiles of H$\alpha$, \ion{He}{1} $\lambda$5867,
{[}\ion{N}{2}{]} $\lambda6583$, and {[}\ion{S}{2}{]} $\lambda$6716
lines (in units of erg cm$^{-2}$ s$^{-1}$ sr$^{-1}$). The intensities
for an O9V star are scaled up by a factor of 2. Bottom panels from
left to right show profiles of line ratios \ion{He}{1}/H$\alpha$,
{[}\ion{N}{2}{]}/H$\alpha$, and {[}\ion{S}{2}{]}/H$\alpha$. Red
and blue lines represent direct and scattered emission lines, respectively.
Thick black lines are total (direct + scattered) intensities or intensity
ratios. Elemental abundances for WNM and hydrogen density of 10 cm$^{-3}$
were assumed for the calculation of the \ion{H}{2} region. The extent
of the dust cloud in the $z$-direction was $|z|\le100$ pc.}
\end{figure*}

Figure \ref{scatt_ratio_O4} shows the dust-scattered halo profiles
of emission lines originating from \ion{H}{2} regions surrounding
O4V and O9V stars, in the homogeneous dusty medium with hydrogen density
of 10 cm$^{-3}$ and the WNM abundances. The top panels of the figures
show brightness profiles of line intensities of H$\alpha$, \ion{He}{1},
{[}\ion{N}{2}{]}, and {[}\ion{S}{2}{]}. Total (scattered + direct),
direct, and scattered intensity profiles are denoted by black, red,
and blue curves, respectively. By comparing the figures, we can confirm
the result that the more extended source (O4V) produces a brighter
dust-scattered halo. The dust-scattered component within the \ion{H}{2}
region surrounding an O4V star is also higher than the less-extended
\ion{H}{2} region surrounding the O9V star. The bottom panels show
the line ratios \ion{He}{1}/H$\alpha$, {[}\ion{N}{2}{]}/H$\alpha$,
and {[}\ion{S}{2}{]}/H$\alpha$. It is clear that the line ratios
{[}\ion{N}{2}{]}/H$\alpha$ and {[}\ion{S}{2}{]}/H$\alpha$ outside
of \ion{H}{2} regions are higher than those at the central regions
and lower than those at ionization boundaries. The \ion{He}{1}/H$\alpha$
profile for the O4V star shows a weak spike at the ionization boundary.
On the other hand, the profile for the O9V star shows a large dip
at the zone where helium is neutral but hydrogen is still ionized.
In the zone, the directly-escaped lines without interaction with dust
grains dominate the line ratio \ion{He}{1}/H$\alpha$ while the dust-scattered
light determines the line ratio outside of the \ion{H}{2} region.
These line ratios found in the dust-scattered halos outside of the
\ion{H}{2} regions match exactly the luminosity ratios found in the
\ion{H}{2} regions, except for a negligible difference in the \ion{He}{1}/H$\alpha$
ratio caused by their wavelength difference. Therefore, we can refer
to the luminosity ratios, for the purpose of estimating the line ratios
in the dust-scattered halos, without detailed calculations of the
dust scattering effect.

\begin{figure*}[t]
\begin{centering}
\includegraphics[scale=0.89]{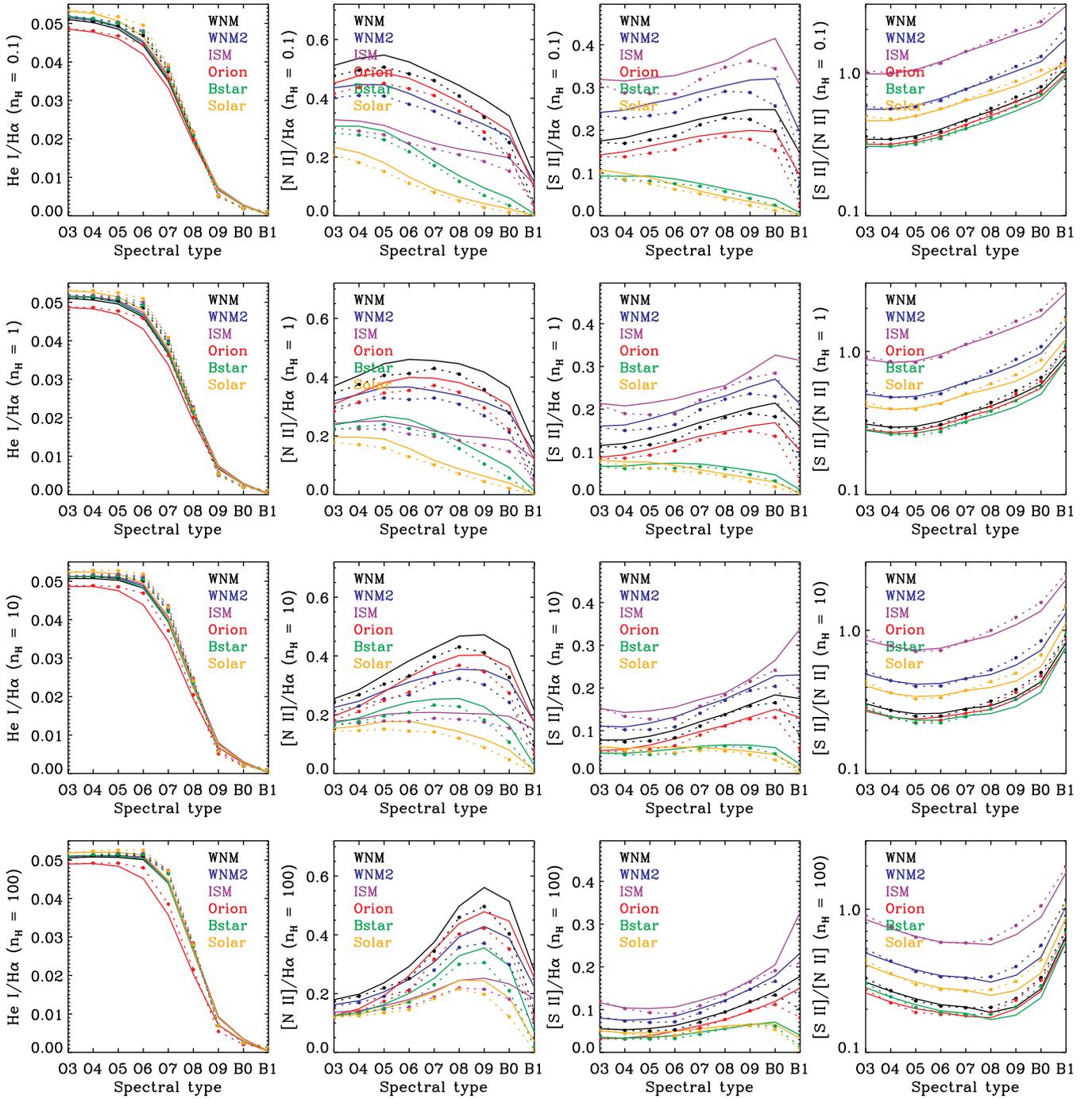}
\par\end{centering}

\caption{\label{luminosity_ratio_cloudy}Luminosity ratios of \ion{He}{1}/H$\alpha$,
{[}\ion{N}{2}{]}/H$\alpha$, {[}\ion{S}{2}{]}/H$\alpha$, and {[}\ion{S}{2}{]}/{[}\ion{N}{2}{]}
for various central ionization sources (from O3V to B1V stars), abundances
(WNM, WNM2, B star, ISM, and Orion nebula) and hydrogen densities
(0.1, 1.0, 10, and 100 cm$^{-3}$). The photoionization models were
calculated with CLOUDY. Solid and dotted lines were obtained by using
the Kurucz and WMBASIC atomospheric models, respectively.}
\end{figure*}

Figure \ref{luminosity_ratio_cloudy} shows the luminosity ratios
of \ion{He}{1}/H$\alpha$, {[}\ion{N}{2}{]}/H$\alpha$, {[}\ion{S}{2}{]}/H$\alpha$,
and {[}\ion{S}{2}{]}/{[}\ion{N}{2}{]} for various hydrogen density,
abundances, and spectral types of ionizing stars. In the figure, the
results obtained not only with the Kurucz (denoted by solid lines),
but also with WMBASIC (dashed lines with dot symbols) stellar models
are shown. The line ratio \ion{He}{1}/H$\alpha$ is quite independent
of elemental abundances and hydrogen density, while {[}\ion{N}{2}{]}/H$\alpha$
and {[}\ion{S}{2}{]}/H$\alpha$ show large variations. Early type
stars always produce higher \ion{He}{1}/H$\alpha$ ratios than late
type stars. The Orion abundances yield slightly lower \ion{He}{1}/H$\alpha$
ratios compared to other abundances, especially for early type stars.
The stellar type giving the highest {[}\ion{N}{2}{]}/H$\alpha$ ratio
varies with hydrogen density, for a given set of abundances. At high
densities, the highest {[}\ion{N}{2}{]}/H$\alpha$ ratio is obtained
from relatively late type stars as compared to at low densities. For
example, for the WNM abundances, the highest value at $n_{{\rm H}}=100$
cm$^{-3}$ is obtained from an O9V star while an O5V star gives the
highest value at $n_{{\rm H}}=0.1$ cm$^{-3}$. Ignoring the B1V star,
the line ratio {[}\ion{S}{2}{]}/H$\alpha$ increases with decreasing
stellar temperature, except for B stars and solar abundances, for
a given set of abundances. We also note that the lowest value of {[}\ion{S}{2}{]}/{[}\ion{N}{2}{]}
has similar dependence on the stellar type as the highest value of
{[}\ion{N}{2}{]}/H$\alpha$ do. At high densities, the lowest value
of {[}\ion{S}{2}{]}/{[}\ion{N}{2}{]} is obtained from relatively
late type stars as compared to at low densities. It is also clear
that abundances affect the line ratios {[}\ion{N}{2}{]}/H$\alpha$
and {[}\ion{S}{2}{]}/H$\alpha$ significantly. The highest {[}\ion{N}{2}{]}/H$\alpha$
and {[}\ion{S}{2}{]}/H$\alpha$ ratios are obtained with the WNM
and ISM abundances, respectively, for a given stellar type. Their
lowest values are generally obtained with the solar abundances. The
abundances of B stars also give relatively low values of {[}\ion{N}{2}{]}/H$\alpha$
and {[}\ion{S}{2}{]}/H$\alpha$. It is also worthwhile noting that
the variations of line ratios due to the difference in the adopted
stellar models is less sensitive than due to the abundances difference.

We should note that the line ratios are intricately tangled with the
adopted abundances. For example, the S abundance of WNM2 is 0.2 dex
larger than that of WNM, while abundances of other elements are the
same. The {[}\ion{S}{2}{]}/H$\alpha$ line ratio obtained from the
WNM2 abundances is then obviously higher than that from the WNM abundances.
However, the {[}\ion{N}{2}{]}/H$\alpha$ line ratio of the WNM abundances
does not remain constant, but becomes lower. Note also that even though
the N abundance between ISM and Orion differs only by 0.05 dex, the
{[}\ion{N}{2}{]}/H$\alpha$ line ratio obtained from the ISM abundances
is lower than that from the Orion abundances by a factor of up to
$\sim2$ because the S abundance of ISM is 0.5 dex larger than that
of Orion. This is because increasing the S abundance changes the heating
and cooling of ionized gas in a complex manner.

We now compare the results with the observed line ratios. The typical
values of {[}\ion{N}{2}{]}/H$\alpha$ and {[}\ion{S}{2}{]}/H$\alpha$
toward local diffuse gas in the Milky Way are $\approx$ 0.3--0.6
and $\approx$ 0.2--0.4, respectively, and they vary substantially
from sightline to sightline. In contrast, classical \ion{H}{2} regions
all cluster near {[}\ion{N}{2}{]}/H$\alpha$ $\approx$ 0.25 and
{[}\ion{S}{2}{]}/H$\alpha$ $\approx$ 0.1. The \ion{He}{1} line
is much fainter (\ion{He}{1}/H$\alpha$ $<$ 0.02) and has been studied
only in a few selected directions. The observed values of {[}\ion{N}{2}{]}/H$\alpha$
and {[}\ion{S}{2}{]}/H$\alpha$ toward local diffuse gas are best
represented by the models with the WNM and WNM2 abundances. Although
early types can reproduce the observed values of {[}\ion{N}{2}{]}/H$\alpha$
at low densities, they cannot reproduce the observed values of {[}\ion{S}{2}{]}/H$\alpha$
and \ion{He}{1}/H$\alpha$. In particular, the observed ratio of
\ion{He}{1}/H$\alpha$ strongly constrains the spectral type of the
ionizing stars to O8 or later. The line ratio of {[}\ion{S}{2}{]}/{[}\ion{N}{2}{]}
$\approx$ 0.5--0.7 is also reasonably well explained with late O
stars and the abundances of WNM and WNM2. We therefore conclude that
late OB stars are the best candidates to match the observed line ratios
of \ion{He}{1}/H$\alpha$, {[}\ion{N}{2}{]}/H$\alpha$, {[}\ion{S}{2}{]}/H$\alpha$,
and {[}\ion{S}{2}{]}/{[}\ion{N}{2}{]}. In addition, our results
indicate that the elemental abundances should be close to those of
WNM and WNM2 to reproduce the observed line ratios. The solar and
B star abundances do not match well the observed line ratios, except
for the case of the highest density ($n_{{\rm H}}=100$ cm$^{-3}$).
The Orion abundances match the observed {[}\ion{N}{2}{]}/H$\alpha$
ratio, but yield slightly weaker {[}\ion{S}{2}{]}/H$\alpha$ ratios
than the observed values. Inversely, the ISM abundances explain the
observed {[}\ion{S}{2}{]}/H$\alpha$, but underpredict the {[}\ion{N}{2}{]}/H$\alpha$
ratio.

\subsection{Three-dimensional Models}

In the previous section, we described the photoionization models in
the homogeneous ISM. However, the real ISM has a filamentary and/or
clumpy structure. A photoionized nebula in a clumpy medium has different
ionic abundances and temperature variation from the models in a uniform
medium \citep{Mathis2005}. The emission coefficients of collisionally
excited lines (CELs), such as {[}\ion{N}{2}{]} and {[}\ion{S}{2}{]},
increase strongly with increasing temperature, while those of RLs,
produced by electron cascades following recombination, are not particularly
temperature sensitive. Therefore, the luminosity ratios between CELs
and the RL H$\alpha$ in clumpy models would be different from those
in uniform models.

To investigate the effect of clumpiness, we first calculated uniform
density models (denoted with a letter ``U'') with the MOCASSIN code.
The hydrogen density of 10 cm$^{-3}$ was assumed for the uniform
models. We next calculated \ion{H}{2} models in a clumpy medium,
assuming a two-phase medium comprising high density clumps embedded
in a low density interclump medium. The volume-averaged density of
hydrogen was also assumed to be 10 cm$^{-3}$, as in the uniform models.
Two values of clumpiness, which can be defined by the density ratio
between dense clumps and interclump medium, were considered. In the
first type of clumpy models (``C1''), dense clumps (cells) have
a hydrogen density of 20 cm$^{-3}$ and occupy a volume fraction of
1/3. The remaining interclump cells have a density of 5 cm$^{-3}$.
In the second type of clumpy models (``C2''), clumps have a density
of 30 cm$^{-3}$ and occupy 1/4 of the total volume. The hydrogen
density of the interclump medium is then assumed to be 10/3 cm$^{-3}$.
The second type is clumpier than the first. For each type of model,
clump cells with high density were randomly cast to occupy the volume
fraction and the remaining cells were then assumed to have lower density.
Spectral types of ionizing source and elemental abundances were varied
as in the one-dimensional models, for each clumpiness (Tables \ref{abundance}
and \ref{sp_type}). We note that \citet{Mathis2005} assumed a hierarchical
density structure for a clumpy medium and compared the resulting ionic
abundances, but not line ratios, with those from the model with uniform
density.

\begin{figure*}[t]
\begin{centering}
\includegraphics[scale=0.89]{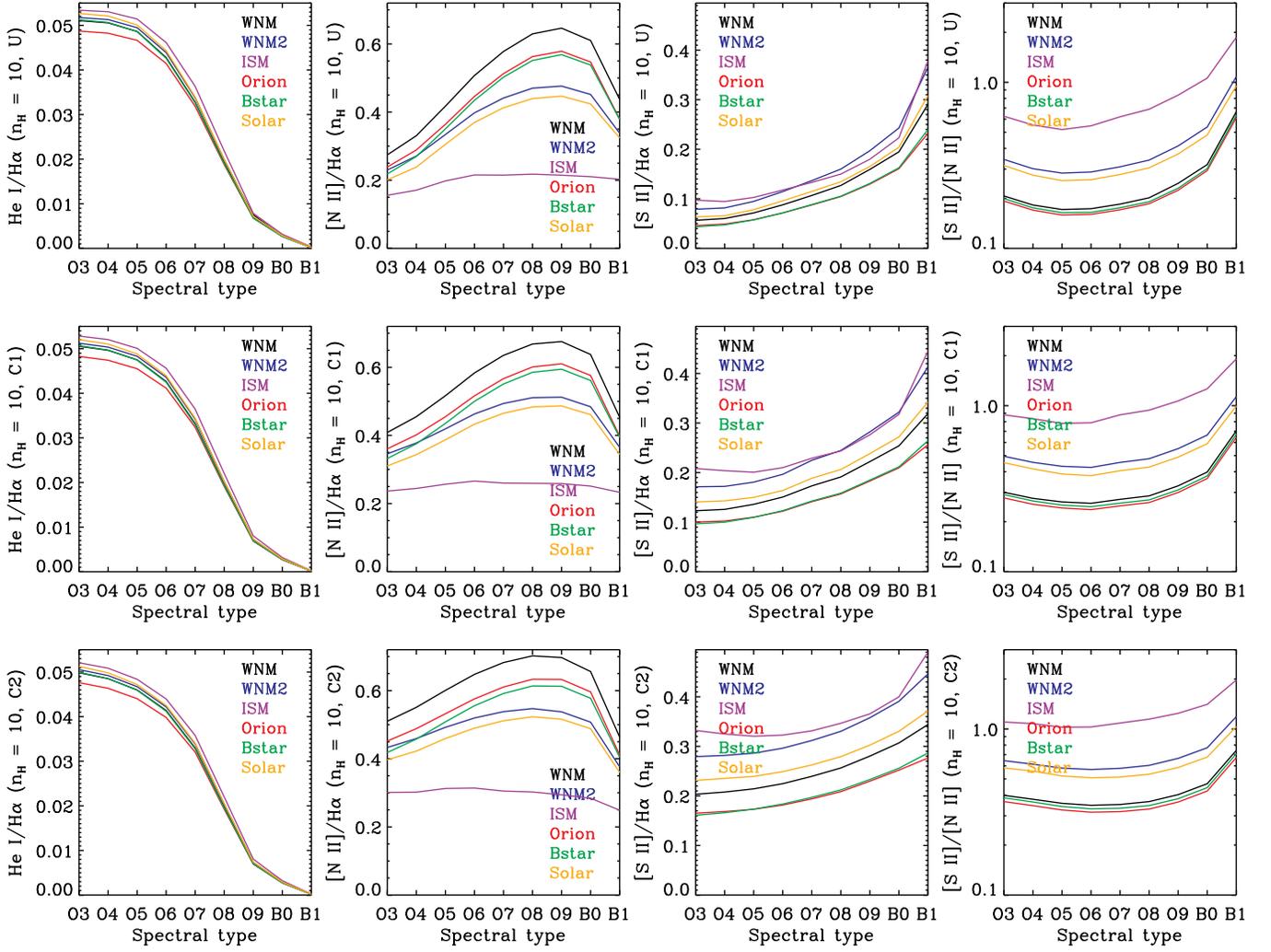}
\par\end{centering}

\caption{\label{luminosity_ratio_mocassin}Luminosity ratios of \ion{He}{1}/H$\alpha$,
{[}\ion{N}{2}{]}/H$\alpha$, {[}\ion{S}{2}{]}/H$\alpha$, and {[}\ion{S}{2}{]}/{[}\ion{N}{2}{]}
for various central ionization sources (from O3V to B1V stars), abundances
(WNM, WNM2, B star, ISM, and Orion nebula). Hydrogen density is 10
cm$^{-3}$. Top, middle, and bottom panels represent the line ratios
obtained from the models ``U,'' ``C1,'' and ``C2,'' respectively.
The models were calculated with MOCASSIN.}
\end{figure*}

Figure \ref{luminosity_ratio_mocassin} summarizes the luminosity
ratios of \ion{He}{1}/H$\alpha$, {[}\ion{N}{2}{]}/H$\alpha$ and
{[}\ion{S}{2}{]}/H$\alpha$ obtained from the uniform and clumpy
models. We first compare the uniform models with the CLOUDY models.
The overall dependences of the line ratios in the uniform model on
the stellar type are generally consistent with the results from the
CLOUDY model. We note, however, that {[}\ion{N}{2}{]}/H$\alpha$
ratios are generally higher than those from the corresponding CLOUDY
models, except for the case of the ISM abundances. For early-type
stars, the {[}\ion{S}{2}{]}/H$\alpha$ ratios are slightly lower
than those obtained from the CLOUDY models. On the other hand, for
the late-type stars, the line ratios {[}\ion{S}{2}{]}/H$\alpha$
are higher than those obtained from the CLOUDY models.

In the figure, the most important result is that the photoionized
\ion{H}{2} regions in the clumpy medium produce higher line ratios
of {[}\ion{N}{2}{]}/H$\alpha$ and {[}\ion{S}{2}{]}/H$\alpha$ than
in the uniform medium. The line ratios {[}\ion{N}{2}{]}/H$\alpha$
and {[}\ion{S}{2}{]}/H$\alpha$ increase as clumpiness increases,
while the change in the \ion{He}{1}/H$\alpha$ ratio is negligible.
This is because dense clumps tend to absorb more H- and He-ionizing
photons than the interclump medium and thus the clumps have higher
temperature than the gas in the uniform models. Luminosities of {[}\ion{N}{2}{]}
and {[}\ion{S}{2}{]} are strongly biased to high density regions
because their emission measures are highly sensitive to temperature
and density. Therefore, their luminosities increase with clumpiness
while the H$\alpha$ and \ion{He}{1} luminosities remain constant
within 2\%. Another interesting result is that the ratios of {[}\ion{N}{2}{]}/H$\alpha$
and {[}\ion{S}{2}{]}/H$\alpha$ increase rather rapidly with clumpiness
for early-type stars, but less rapidly for late-type stars. The {[}\ion{N}{2}{]}/H$\alpha$
ratio of the ``C2'' models is larger than that of the uniform models
by a factor of $\sim$1.9--2.0 for O3V star and a factor of $\sim$1.1--1.3
for O9V star, depending on the adopted abundances. The {[}\ion{S}{2}{]}/H$\alpha$
ratio is increased by a factor of $\sim$2.4--2.7 for O3V star and
a factor of $\sim$1.8--2.0 for O9V star. It is also worth noting
that the {[}\ion{S}{2}{]}/{[}\ion{N}{2}{]} ratio becomes less dependent
on the stellar type as clumpiness increases.

Note that the line ratios of {[}\ion{N}{2}{]}/H$\alpha$, {[}\ion{S}{2}{]}/H$\alpha$
and {[}\ion{S}{2}{]}/{[}\ion{N}{2}{]} depend strongly on the elemental
abundances. The line ratio of {[}\ion{N}{2}{]}/H$\alpha$ obtained
from the ``C2'' models, except those with the ISM abundances, matches
well the observed ratio in the diffuse ISM, regardless of the stellar
type. In particular, a {[}\ion{N}{2}{]}/H$\alpha$ line ratio of
up to $\sim0.7$ is predicted for the WNM abundances and O9V star.
However, the stellar type of the ionizing source is strongly constrained
to O8V or later-type stars by the observed ratio of \ion{He}{1}/H$\alpha$,
as in the CLOUDY models.

The differences of the {[}\ion{N}{2}{]}/H$\alpha$ and {[}\ion{S}{2}{]}/H$\alpha$
line ratios in the MOCASSIN models from those of the CLOUDY models
might be due to atomic data differences adopted in the codes. For
example, the data set for collision strengths, transition probabilities,
and energy levels was adopted from version 5.2 of the CHIANTI atomic
database \citep{Landi2006}. However, the CHIANTI database is not
included by default in calculations with the CLOUDY code. We also
note that reliable data for the recombination rate coefficients for
S are not available, except S$^{2+}$. The fact that the models with
the WNM and WNM2 abundances produced significantly different {[}\ion{N}{2}{]}/H$\alpha$
ratios demonstrates the complicated relationship between the atomic
species. Using the WNM2 abundances in the MOCASSIN models, we also
obtained the same trend as in the case of the CLOUDY code. Therefore,
judging which models are more reliable over others without reliable
atomic data sets would be impractical. However, it is true that clumpy
models produce higher line ratios than uniform models. The clumpy
models considered in the present study demonstrate the effect of clumpiness.
A more realistic clumpy medium, such as a hierarchical density structure,
might help to understand the effect of clumpiness in photoionization
models.

\begin{figure*}[t]
\begin{centering}
\includegraphics[scale=0.8]{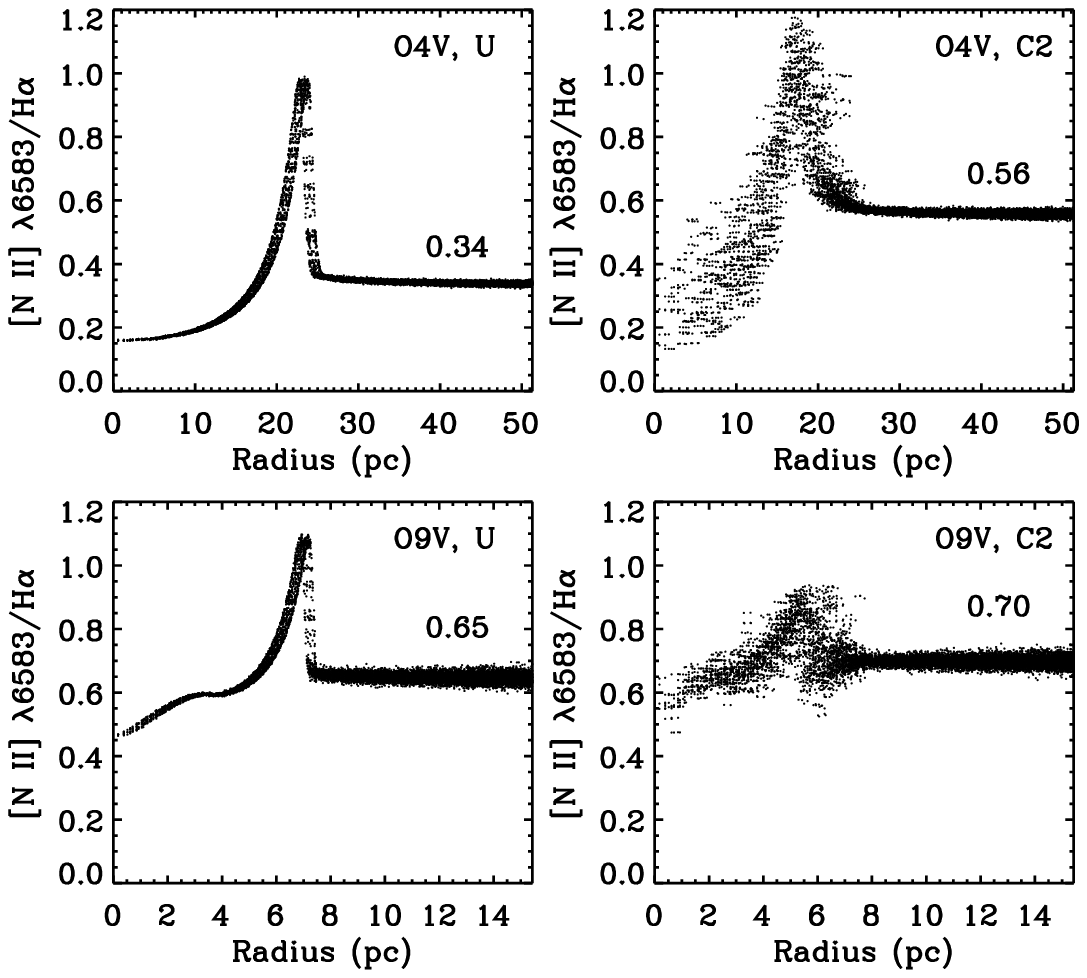}\ \ \ \includegraphics[scale=0.8]{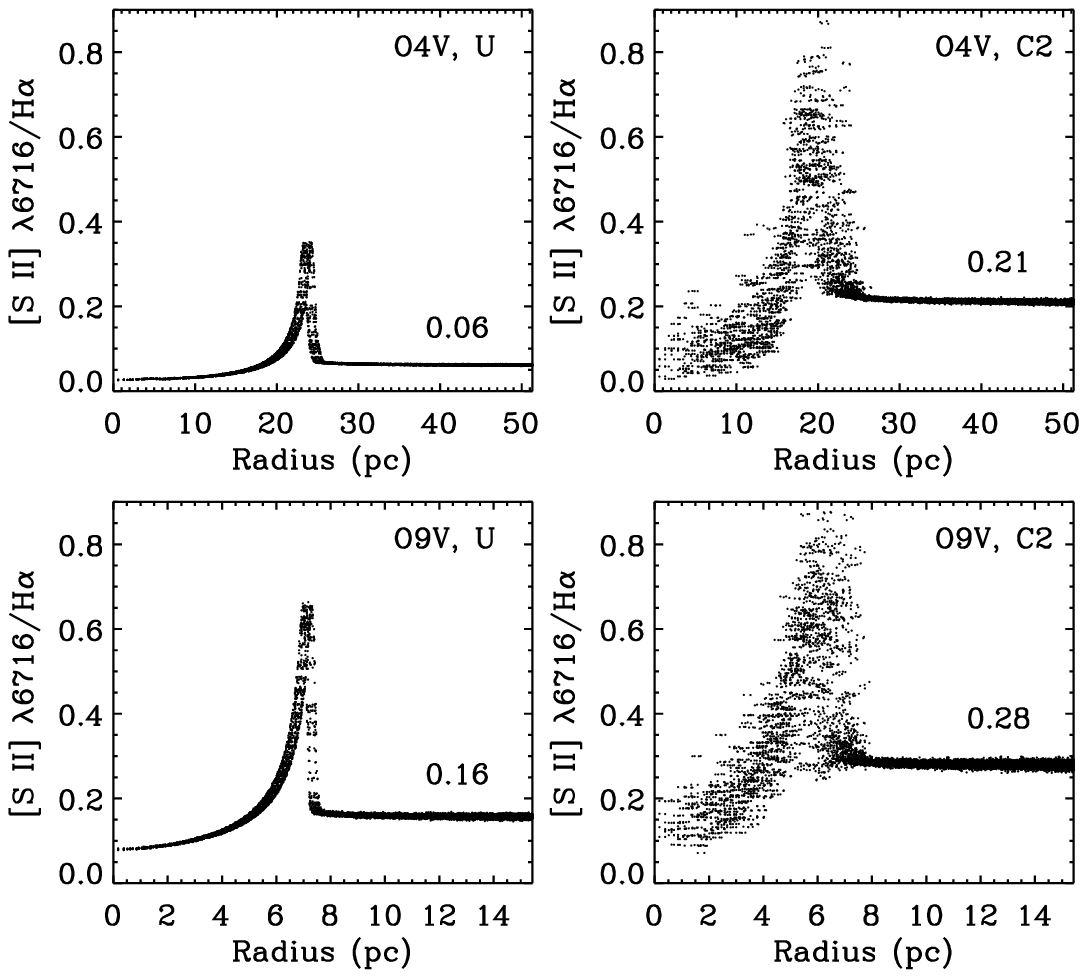}
\par\end{centering}

\caption{\label{scat_moc_NII_SII}Line ratios ({[}\ion{N}{2}{]}/H$\alpha$,
and {[}\ion{S}{2}{]}/H$\alpha$) in the dust-scattered halos of \ion{H}{2}
regions surrounding O4V and O9V stars. Average hydrogen density of
10 cm$^{-3}$ and the WNM abundances were assumed. The left four figures
represent line ratios of {[}\ion{N}{2}{]}/H$\alpha$, and the right
four figures {[}\ion{S}{2}{]}/H$\alpha$. Numbers indicate the intensity
ratios at the dust-scattered halo outside of \ion{H}{2} regions.}
\end{figure*}

We now examine the detailed profiles of the dust-scattered halos surrounding
\ion{H}{2} regions calculated with the clumpy models. For the dust
radiative transfer models, we assumed a dusty cloud with a constant
density of $n_{{\rm H}}=10$ cm$^{-3}$. The physical dimension of
the cloud was the same as in the CLOUDY models. For the clumpy medium
models, we may need to use the same clumpy density distribution as
used in the photoionization models. However, because the optical depths
within the \ion{H}{2} regions are small and the clumpiness in the
dust cloud does not affect the line ratios, we ignored the clumpiness
in the dust cloud. Figure \ref{scat_moc_NII_SII} shows the brightness
profiles of the {[}\ion{N}{2}{]}/H$\alpha$ and {[}\ion{S}{2}{]}/H$\alpha$
line ratios in the \ion{H}{2} regions produced by O4V and O9V stars
and their dust-scattered halos. The results were obtained from the
uniform and clumpy ``C2'' models with the WNM abundances. It is obvious
that both line ratios are increased in the clumpy models. However,
the maximum ratios show somewhat different behavior. The maximum value
of {[}\ion{N}{2}{]}/H$\alpha$ near the ionization boundary increased
with clumpiness for early-type stars, such as O3V, while the maximum
value decreased for late-type stars. On the other hand, the maximum
value of {[}\ion{S}{2}{]}/H$\alpha$ always increased with clumpiness,
regardless of stellar type.

This property may be largely related to the fate of He-ionizing photons.
As noted in \citet{Mathis2005}, the most important factor to determine
the ionization states in a clumped density distribution is the fate
of He-ionizing radiation, depending on the hardness of the stellar
radiation. Hot early-type stars produce such plentiful He-ionizing
photons that the He ionization zone coincides with the H ionization
zone in a uniform medium. The effect of the clumpy medium is to decrease
the ionic fraction (He$^{+}$/He) and reduce the size of the He ionization
zone. The ionic fraction (N$^{2+}$/N) follows a similar pattern as
He$^{+}$. Therefore, the ionic fraction (N$^{+}$/N) increases near
the ionization front and thus increases the maximum value of {[}\ion{N}{2}{]}/H$\alpha$.
However, the He ionization zone for cool later-type stars is smaller
than the H ionization zone. In this case, the He-ionizing photons
tend to penetrate dilute interclump regions beyond the He ionization
zone that is expected from uniform models. The same effect is applied
to N$^{2+}$, and thus (N$^{+}$/N) becomes a bit lower at the ionization
boundary compared to the uniform models. On the other hand, since
the ionization potential for S$^{+}$ $\rightarrow$ S$^{2+}$ is
smaller than the He-ionizing potential and the ionization potential
for S$^{0}$ $\rightarrow$ S$^{+}$ is smaller than the H-ionizing
potential, the total ionic fraction of S$^{+}$ and the maximum value
of {[}\ion{S}{2}{]}/H$\alpha$ always increase with clumpiness, regardless
of the stellar type of the ionization source.

\begin{figure*}[!]
\begin{centering}
\includegraphics[scale=0.9]{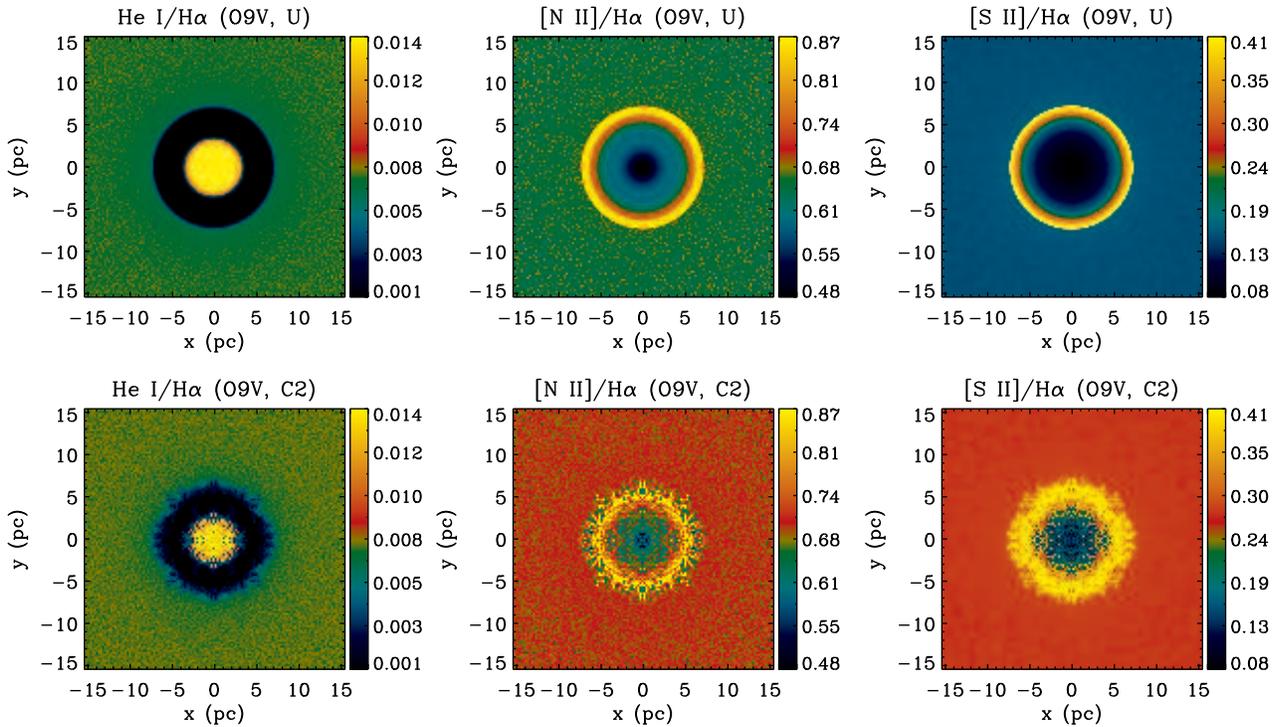}
\par\end{centering}

\caption{\label{scat_moc_O9V10C2}Comparison between the line ratio maps (\ion{He}{1}/H$\alpha$,
{[}\ion{N}{2}{]}/H$\alpha$, and {[}\ion{S}{2}{]}/H$\alpha$ from
left to right) obtained from the uniform (top) and clumpy (bottom)
models.}
\end{figure*}

Figure \ref{scat_moc_O9V10C2} shows sample images of the line ratios
calculated for an O9V star to demonstrate the difference between the
uniform and clumpy models. It is clear that transition regions in
the line ratio maps near the ionization boundary become less sharp
for the clumpy models. In some cases, the clumpiness may produce a
smooth transition zone and the line ratios would smoothly change to
the background values, as in the model of the \ion{H}{2} region surrounding
the O9.5 V star $\zeta$ Oph \citep{Wood2005}.

\section{Effect of Stellar H$\alpha$ Absorption Line}

In the previous section, we reproduced the typical values of the {[}\ion{N}{2}{]}/H$\alpha$
and {[}\ion{S}{2}{]}/H$\alpha$ line ratios in the diffuse ISM regions
by photoionization due to late OB stars. However, the highest line
ratios may require additional non-ionizing heating sources, which
will be discussed in Section 5. We here consider another possibility
to explain the elevated {[}\ion{N}{2}{]}/H$\alpha$ and {[}\ion{S}{2}{]}/H$\alpha$
line ratios in the faint H$\alpha$ regions, which has never been
discussed sufficiently in analyzing the diffuse optical emissions.

\subsection{Implications from Integrated Spectra of Galaxies}

We should note that most, if not all, of the spiral galaxies, including
the Milky Way, showing the diffuse H$\alpha$ emission outside of
bright \ion{H}{2} regions are Sb or later types, including irregulars
(e.g., \citealp{Rossa2003,Hunter1990,Kennicutt1995}), in which the
underlying stellar continuum is dominated by B- and A-type stars \citep{Kennicutt1992a,Kennicutt1992b}.
Balmer absorption lines in the underlying stellar continuum can give
rise to underestimation of the diffuse H$\alpha$ intensity and overestimation
of the line ratios of forbidden lines, such as {[}\ion{N}{2}{]} and
{[}\ion{S}{2}{]}, over H$\alpha$. The equivalent widths (EWs) of
the Balmer absorption line as a function of stellar spectral type
are given by \citet{Pickles1998}. The maximum H$\alpha$ absorption
EW of $\sim10$\AA\ occurs at early A-type stars. \citet{Bica1986}
shows the EW of the H$\alpha$ absorption line as a function of age
in an evolving stellar population. For a Sb or Sc disk galaxy with
a mixed population, it appears that an H$\alpha$ absorption EW of
$\approx4-5$\AA\ should be expected for the background faint star
population outside of OB associations. The EWs of nebular emission
lines in the integrated spectra of Sb and Sc galaxies are $\sim5-100$\AA\
\citep{Kennicutt1992a,Kennicutt1992b,Sodre1999}. Therefore, the underlying
H$\alpha$ absorption line would appreciably increase the line ratios
{[}\ion{N}{2}{]}/H$\alpha$ and {[}\ion{S}{2}{]}/H$\alpha$ not
only in the integrated spectra, but also in the diffuse ISM regions.
For example, for galaxies with the emission EW of $\sim10$\AA, the
line ratios will be increased by a factor of $\sim$2. The effect
of H$\alpha$ absorption becomes less important for the galaxies with
older stellar population. However, even the oldest stellar population
with an age of $\gtrsim13$ Gyr has an absorption EW of $\gtrsim1$\AA\
\citep{Bica1986}.

Since about half of the H$\alpha$ emission is known to originate
from the diffuse ISM regions (DIG) and the mean ratios of {[}\ion{N}{2}{]}/H$\alpha$
and {[}\ion{S}{2}{]}/H$\alpha$ in the DIG is higher than those in
\ion{H}{2} regions by a factor of $\approx2$, the line ratios in
the integrated spectrum over an entire galaxy are expected to be higher
than the mean values of \ion{H}{2} regions by a factor of $\approx1.5$.
We therefore expect that the line ratios of integrated spectra of
star forming galaxies, after correction of the underlying stellar
absorptions, would be significantly different from those of individual
\ion{H}{2} regions in the diagnostic line-ratio diagrams \citep[i.e., the BPT diagram; ][]{Baldwin1981,Veilleux1987},
if the enhancement of the line ratios in the DIG were not due to the
underlying Balmer absorption lines.

\citet{Lehnert1994} used the integrated optical spectra of galaxies,
published by \citet{Kennicutt1992a,Kennicutt1992b}, with H$\alpha$
emission-line EWs $>$ 30\AA\ in order to avoid errors in the measured
line ratios caused by underlying stellar absorption lines. They corrected
the measured line ratios assuming that both the H$\alpha$ and H$\beta$
stellar absorption lines have EWs of 5\AA\ and found that in general
the line ratios in the integrated galaxy spectra occupy a region that
is intermediate between \ion{H}{2} regions and the AGNs. The resulting
{[}\ion{S}{2}{]} $\lambda\lambda$6716, 6731/H$\alpha$ ratios, for
a given {[}\ion{O}{3}{]} $\lambda$5007/H$\beta$, were typically
a factor of $\sim1.5$ larger, while the {[}\ion{N}{2}{]} $\lambda$6783/H$\alpha$
ratios were more similar to those of \ion{H}{2} regions, only being
typically enhanced by $\sim25$\%, which is not very significant and
well within the scatter expected for individual \ion{H}{2} regions.
They therefore concluded that at least 25\% of the observed H$\alpha$
flux in their sample galaxies would arise in the DIG.

However, more recent results seem to indicate no significant difference
between individual \ion{H}{2} regions and star-forming galaxies in
the diagnostic diagrams, after careful subtraction of the underlying
Balmer absorption lines. Using the same dataset as \citet{Lehnert1994},
\citet{Sodre1999} made a better correction for the stellar absorption
EWs by correlating the observed H$\alpha$ and H$\beta$ EWs. They
found that the diagnostic diagram of integrated ratios {[}\ion{O}{3}{]}/H$\beta$
versus {[}\ion{N}{2}{]}/H$\alpha$ follows the sequence of the giant
\ion{H}{2} regions very well. In their diagram of {[}\ion{O}{3}{]}/H$\beta$
versus {[}\ion{S}{2}{]}/H$\alpha$, the absorption-corrected ratios
are also well within the \ion{H}{2} region sequence (except that
those with {[}\ion{O}{3}{]}/H$\beta$ $>1$ lie at the upper boundary
defined by individual \ion{H}{2} regions). \citet{Moustakasetal2006}
used a spectral synthesis fitting code to find the stellar model continuum,
including Balmer absorption lines, that optimally reproduces the integrated
continuum spectrum and estimated pure nebular emission line EWs of
412 star-forming galaxies after subtracting the best-fit continuum
spectra. They noted that the sequences formed by individual \ion{H}{2}
regions and star-forming galaxies overlap across the full range of
emission-line ratios in the {[}\ion{O}{3}{]} $\lambda$5007/H$\beta$
versus {[}\ion{N}{2}{]}/H$\alpha$ plane. However, in the {[}\ion{N}{2}{]}/H$\alpha$
versus {[}\ion{S}{2}{]}/H$\alpha$ plane (Fig.~7 of \citealp{Moustakas_Kennicutt2006}),
the absorption-corrected line ratios preferentially lie at or slightly
outside the envelope defined by \ion{H}{2} regions. We note that
this cannot be caused by a relative enhancement or reduction of the
integrated {[}\ion{N}{2}{]}/H$\alpha$ ratios compared to the ratios
of \ion{H}{2} regions, but can be caused by an enhancement of the
{[}\ion{S}{2}{]}/H$\alpha$ ratios in integrated spectra by about
0.1 dex. We also note that Fig.~1 of \citet{Brinchmann2004}, which
was obtained by fitting the stellar synthesis model continuum to the
Sloan Digital Sky Survey data, shows that star-forming galaxies lie
at the same locations as \ion{H}{2} regions in the diagnostic diagram
of {[}\ion{O}{3}{]}/H$\beta$ versus {[}\ion{N}{2}{]}/H$\alpha$.

Unlike the absorption-corrected line ratios in the integrated galaxy
spectra, the line ratios observed in the DIG are significantly different
from individual \ion{H}{2} regions in the diagnostic diagrams (see
Fig.~7 of \citet{Moustakas_Kennicutt2006} and Fig.~2(a) of \citet{Flores-Fajardo2009}).
Therefore, the enhancement of {[}\ion{N}{2}{]}/H$\alpha$ in the
diffuse ISM regions may be due mainly to the underlying absorption
lines. In other words, the ratio of {[}\ion{N}{2}{]}/H$\alpha$ in
the diffuse ISM or DIG regions is not likely to be significantly different
from that of \ion{H}{2} regions if the underlying stellar absorption
lines are properly appreciated. In the case of {[}\ion{S}{2}{]}/H$\alpha$,
the line ratios observed outside \ion{H}{2} regions, after correction
of the stellar absorption line, may be slightly larger than that of
\ion{H}{2} regions. However, the enhancement would be only by a factor
of $\approx1.5$, taking into account the facts that half of the optical
lines originate from the diffuse regions and the ratio in integrated
spectra is only 0.1 dex higher than that of \ion{H}{2} regions. It
is also worth noting that the {[}\ion{S}{2}{]} line arises not only
in the H$^{+}$/H$^{0}$ interface regions of ionized nebula, but
also in the H$^{0}$--C$^{+}$ regions of photodissociation regions
(PDRs) illuminated by strong FUV radiation, because the ionization
potential of S$^{0}$ is low (10.36 eV) \citep{Petuchowski1995,Storzer2000}.
\citet{Storzer2000} found that the relative strength of PDR emission
to H$^{+}$/H$^{0}$ interface emission is $\approx0.14-1.5$. Therefore,
in a certain condition with high density and strong FUV radiation,
the PDR may contribute to the {[}S II{]} flux in integrated spectra
of star forming galaxies. Taking the contribution of PDR emission
into account, the difference in the {[}\ion{S}{2}{]}/H$\alpha$ line
ratio between the DIG and \ion{H}{2} regions would be reduced even
further. However, it is not certain, at this point, how significantly
the PDR emission would contribute to the integrated spectra of star
forming galaxies. The {[}S II{]} flux from PDRs may yield a spatially
dependent contribution to the {[}\ion{S}{2}{]}/H$\alpha$ line ratio
because molecular clouds and PDRs are rarer, especially, in interarm
regions.

We also note that some authors assumed the same EW for all Balmer
absorption lines (H$\alpha$, H$\beta$, H$\gamma$, and H$\delta$)
\citep[e.g., ][]{Lehnert1994,LeeJC_KISS_2004}. We obtained EW(H$\beta$)
= 1.52 EW(H$\alpha$), EW(H$\gamma$) = 1.38 EW(H$\alpha$), and EW(H$\delta$)
= 1.36 EW(H$\alpha$) by fitting the stellar spectral data in \citet{Pickles1998}.
Therefore, the EWs of H$\beta$, H$\gamma$, and H$\delta$ could
be assumed to be the same. However, assuming the same EW for H$\alpha$
and H$\beta$ would cause not only an error in dust extinction correction,
but also a shift in the diagnostic diagrams, such as {[}\ion{O}{3}{]}
$\lambda$5007/H$\beta$ versus {[}\ion{N}{2}{]}/H$\alpha$.

In summary, it is likely that the line ratios observed in the DIG
are not significantly different from those of \ion{H}{2} regions.
In the following section, we will further demonstrate how the stellar
absorption lines are able to increase substantially the line ratios
in the diffuse ISM regions.

\subsection{Effect on the Lines Ratios in the Diffuse ISM}

If significant variations occur in stellar populations vertically
as well as across spiral arms, the faint H$\alpha$ regions would
be significantly affected by the absorption feature in the underlying
continuum. OB stars are usually concentrated at star forming regions
in spiral arms, while A- and later-type stars are relatively widespread.
The scale-height of late-type stars is also higher than those of earlier
types. Therefore, not only the direct starlight, but also the scattered
light from late-type stars would be much more extended than those
from OB stars. In our Galaxy, for example, the sky can be divided
into places where OB stars or A-type stars are dominant, as clearly
shown in Figure 15 of \citet{Seon2011a}. Within and near \ion{H}{2}
regions, the stellar continuum would be dominated by OB stars and
the H$\alpha$ emission line would overwhelm the underlying H$\alpha$
absorption feature in the continuum component from later-type stars.
At locations far from OB stars, where later-type stars dominate the
continuum, the H$\alpha$ emission line will rapidly decrease while
the underlying H$\alpha$ absorption line is far more uniform. Therefore,
the absorption feature would carve out the emission line. The effect
of the H$\alpha$ absorption feature in the continuum would gradually
increase with the distance from bright \ion{H}{2} regions and with
a decrease of the H$\alpha$ intensity. Therefore, the stellar absorption
feature should significantly influence faint areas where the continuum
is dominated by A- or later-type stars.

\begin{figure*}[t]
\begin{centering}
\includegraphics[scale=0.5]{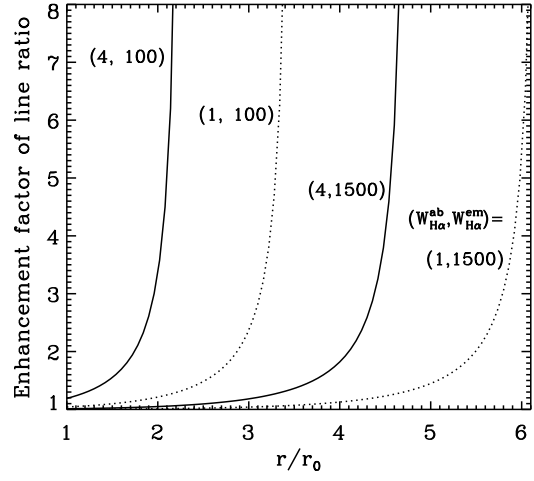}
\par\end{centering}

\caption{\label{fig:ratio_increase}Increasing factor of the line ratios {[}\ion{N}{2}{]}/H$\alpha$
and {[}\ion{S}{2}{]}/H$\alpha$, due to the underlying H$\alpha$
absorption, as a function of distance from an \ion{H}{2} region.
Here, $r_{0}(\approx l_{{\rm mfp}})$ is the distance from which Equation
\ref{eq:halo_profile} is applicable. EW of H$\alpha$ absorption
line relative to the diffuse continuum and EW of H$\alpha$ emission
line relative to continuum of OB association are denoted in parenthesis
($W_{{\rm H}\alpha}^{{\rm ab}}$, $W_{{\rm H}\alpha}^{{\rm em}}$).}
\end{figure*}

We can quantitatively demonstrate how the underlying H$\alpha$ absorption
line can significantly affect the measured line ratios, if the diffuse
optical emission lines are mostly caused by dust scattering of photons
originating from bright \ion{H}{2} regions. First of all, we assume
that the surface brightness of the dust-scattered continuum from an
OB association $I_{{\rm cont}}^{{\rm OB}}(r)$ decreases as in Equation
\ref{eq:halo_profile}, while the diffuse continuum background $I_{{\rm cont}}^{{\rm late}}$
surrounding the OB association is dominated by late-type stars and
thus spatially uniform. The observed line ratio {[}\ion{N}{2}{]}/H$\alpha$
(or {[}\ion{S}{2}{]}/H$\alpha$) at a distance $r$ (from the OB
association) larger than the mean free path ($r_{0}\approx l_{{\rm mfp}}$)
in the dust-scattered halo of the OB association can be related to
the real emission line ratio ($I_{{\rm [NII]}}^{{\rm em}}/I_{{\rm H}\alpha}^{{\rm em}}$)
by 
\begin{eqnarray}
\frac{I_{{\rm [NII]}}^{{\rm obs}}(r)}{I_{{\rm H}\alpha}^{{\rm obs}}(r)} & = & \frac{I_{{\rm [NII]}}^{{\rm em}}}{I_{{\rm H}\alpha}^{{\rm em}}}\left[1-\frac{W_{{\rm H}\alpha}^{{\rm ab}}I_{{\rm cont}}^{{\rm late}}}{W_{{\rm H}\alpha}^{{\rm em}}I_{{\rm cont}}^{{\rm OB}}(r)}\right]^{-1}\label{eq:ratio_halo}\\
 & = & \frac{I_{{\rm [NII]}}^{{\rm em}}}{I_{{\rm H}\alpha}^{{\rm em}}}\left[1-\frac{W_{{\rm H}\alpha}^{{\rm ab}}I_{{\rm cont}}^{{\rm late}}}{W_{{\rm H}\alpha}^{{\rm em}}I_{{\rm cont}}^{{\rm OB}}(r_{0})}\left(\frac{r}{r_{0}}\right)^{2}e^{(r_{0}/r_{{\rm c}})(r/r_{0}-1)}\right]^{-1},\nonumber 
\end{eqnarray}
where $W_{{\rm H}\alpha}^{{\rm ab}}$ and $W_{{\rm H}\alpha}^{{\rm em}}$
denote the H$\alpha$ absorption EW measured relative to the diffuse
continuum background ($I_{{\rm cont}}^{{\rm late}}$) and the emission
EW relative to the continuum of the OB association ($I_{{\rm cont}}^{{\rm OB}}(r)$),
respectively. In the dust-scattered halo, the EWs are constant. The
emission line ratio $I_{{\rm [NII]}}^{{\rm em}}/I_{{\rm H}\alpha}^{{\rm em}}$
will be constant in the dust-scattered halo, while the line intensities
decrease together with the continuum intensity from the OB association
as distance increases. In general, $W_{{\rm H}\alpha}^{{\rm em}}$
of bright \ion{H}{2} regions in disc galaxies ranges from 100\AA\
to 1500\AA, with a mean value of 400\AA\ \citep{Bresolin1997}.
As noted earlier, the most probable absorption EW would be $W_{{\rm H}\alpha}^{{\rm ab}}\approx4$\AA\
\citep{Bica1986}. We consider, however, not only the most probable
$W_{{\rm H}\alpha}^{{\rm ab}}=4$\AA\, but also the worst case of
$W_{{\rm H}\alpha}^{{\rm ab}}=1$\AA. From Figures \ref{ha_halo_profile2}
and \ref{Rband}, we can assume, for $n_{{\rm H}}=10$ cm$^{-3}$,
that $I_{{\rm cont}}^{{\rm late}}/I_{{\rm cont}}^{{\rm OB}}(r_{0})\approx4$
at $r_{0}\approx l_{{\rm {\rm mfp}}}$ = 57 pc and $r_{0}/r_{{\rm c}}=56.9/72.4=0.79$.
Figure \ref{fig:ratio_increase} shows the resulting enhancement factor
of the line ratio as a function of distance for various combinations
of ($W_{{\rm H}\alpha}^{{\rm ab}}$, $W_{{\rm H}\alpha}^{{\rm em}}$).
As shown in the figure, the line ratio increases significantly within
a few hundred pc. The enhancement factor increases at closer distance
when the absorption EW is higher and/or the emission EW is lower.

Note that the forbidden-to-H$\alpha$ line ratios, especially {[}\ion{S}{2}{]}/H$\alpha$,
are observed to be enhanced by a factor of up to 8 (or more) compared
to the ratios in \ion{H}{2} regions (e.g., Fig.~12 in \citealp{Madsen06}).
However, such a large increase by a factor of $\gtrsim3$ seems to
be rare. In this regard, it is important to note that the line ratio
increases relatively slowly up to a factor of $\sim3$ and then rapidly
beyond the point. Therefore, the enhancement factor $\gtrsim3$ would
be found only in a small outermost part of the halo and be rarely
observed. In addition, we assumed only a single \ion{H}{2} region
and its scattered halo. In reality, there should be other \ion{H}{2}
regions. The emission-lines (H$\alpha$, {[}\ion{N}{2}{]} and {[}\ion{S}{2}{]})
originating from the neighboring \ion{H}{2} regions will be scattered
to the halo under consideration. The scattered H$\alpha$ from other
\ion{H}{2} regions may soften the H$\alpha$ absorption effect and
the rapid increase of line ratios at the outermost part of the halo.

In the Wisconsin H$\alpha$ Mapper (WHAM) survey of the Milky Way,
the bright stars with $m_{V}<5.5$ were flagged to denote possible
contamination due to strong H$\alpha$ absorption features \citep{Haffner2003}.
However, there still remains dust-scattered starlight called DGL \citep{Toller1987,Brandt2012}
and light from stars with $m_{V}>5.5$. The average EW of the diffuse
H$\alpha$ background of our Galaxy is estimated to be 11\AA\ \citep{Brandt2012},
assuming the interstellar radiation field of \citet{Mathis1983}.
In the external galaxies, the H$\alpha$ absorption feature in the
underlying continuum both from unresolved stars and the diffuse dust-scattered
starlight would hamper the estimation of the line ratios. To the best
of our knowledge, \citet{Walterbos1994} was the only one in which
the effect of underlying Balmer absorption lines was discussed in
investigating line ratios in the diffuse ISM. For M\,31, the integrated
H$\alpha$ flux corresponds to an EW of only 4--6 \AA\ \citep{Walterbos1994}.
Therefore, the underlying H$\alpha$ absorption line is expected to
strongly affect the observed line ratios in the diffuse ISM regions.
Broadband surface photometry of M\,31 shows typical colors of $B-V=0.6-0.8$,
and $U-B=0.0-0.25$, which are different from a pure A-star spectrum.
This led \citet{Walterbos1994} to claim that the effect due to the
H$\alpha$ absorption line are not likely to be important. They also
performed a simple test, to reduce the effect of underlying H$\alpha$
absorption line, by subtracting a 10\% lower continuum from the H$\alpha$
filter image and found no evidence on the effect in the H$\alpha$
map. However, as noted by them, a more detailed test requires high-resolution
multicolor imaging, combined with spectroscopy to constrain the stellar
population and its small scale variations.

\section{Discussion}

In this section, we present additional support for our results and
argue the necessity of reconsidering the dust-scattering origin of
the diffuse H$\alpha$ emission more seriously. In fact, we should
note that there is no direct evidence, given the results presented
in the previous sections, supporting the contention that most of the
diffuse H$\alpha$ emission originates from in situ ionized gas. The
previous studies of \citet{Mattila2007}, \citet{Witt2010}, and \citet{Seon2011b}
have mainly dealt with the dust-scattered H$\alpha$ photons at high
Galactic latitudes. The present results strongly suggest the significance
of dust scattering in the diffuse H$\alpha$ emission not only at
high-latitude clouds but also near \ion{H}{2} regions.

\subsection{Dust-Scattered H$\alpha$ Halos}

There is observational evidence that at least some parts of the diffuse
H$\alpha$ emission outside of classical bright \ion{H}{2} regions
may be in fact caused by dust scattering. If the dust scattering is
the dominant source of the diffuse H$\alpha$ halos around the bright
\ion{H}{2} regions, the halos would not be observable or would be
very faint in a radio recombination continuum that is almost free
from dust-scattering. In this way, at least some part of the extended
H$\alpha$ halo of the Orion \ion{H}{2} region was attributed to
the dust-scattered halo \citep{Subrahmanyan2001,ODell2009}. More
direct evidence of dust-scattered origin of the H$\alpha$ halo can
be provided by observations of the H$\alpha$ polarization. \citet{Topasna1999}
attempted to detect H$\alpha$ polarizations by selective extinction
in the Monoceros supernova remnant, the Rosette Nebula, and the North
America Nebula. While these were not detected, H$\alpha$ polarizations
by scattering in dust shells around the Rosette Nebula and the North
America Nebula were instead observed.

Additional support for the importance of dust scattering in the diffuse
H$\alpha$ emission is provided from the large scale surveys of our
Galaxy. \citet{Kutyrev2001,Kutyrev2004} carried out a pilot survey
of the Galactic plane with 2.17 $\mu$m Br$\gamma$ hydrogen RL, which
is relatively free from dust-scattering, and found a typical filling
factor of a Br$\gamma$-emitting gas of $\sim1$ \%, which is much
smaller than the value derived for H$\alpha$. Recent observations
using the Wilkinson Microwave Anisotropy Probe have found that the
ratio of the free\textendash{}free radio continuum to H\textgreek{a}
is surprisingly low in the WIM \citep{Davies2006,Dobler2008a,Dobler2008b,Dobler2009,Gold2011}.
These discrepancies likely suggest the significant role of dust scattering
in producing the diffuse H$\alpha$ emission outside of bright \ion{H}{2}
regions. However, we note that this discrepancy is not just present
in regions where dust scattering may contribute, but even in regions
that appear to be rather dust free, e.g., the Gum nebula. The cause
of this observed fact may therefore be more complex and involve other
factors besides scattering.

It is also of interest to note that a correlation between H$\alpha$
and the anomalous dust emission at $\sim30$ GHz, which is attributed
to spinning very small dust grains, is found in the WMAP data \citep{Dobler2008a,Dobler2008b}.
The spectrum of H$\alpha$-correlated microwave emission does not
follow the expected free-free spectrum from the WIM, but the electric
dipole emission from rapidly rotating dust grains. Since then, there
have been a number of studies showing the association of peaks in
the anomalous dust emission with individual interstellar clouds \citep[e.g., ][]{Vidal2011}.
The spinning dust microwave emission is simply the signal from the
smallest component of the grain size distribution in interstellar
space. The correlation of the anomalous dust emission with H$\alpha$
is explained most readily, if a significant fraction of the H$\alpha$
is also coming from dust via scattering.

The photoionization models of the diffuse H$\alpha$ emission assume
no significant absorption of the Lyc \citep{Domgorgen94}. A widely
accepted view for the almost-free escape of Lyc from the bright \ion{H}{2}
regions is to assume a highly clumpy or turbulent density structure
of ISM with a lot of almost-empty space. \citet{Ferguson96a,Ferguson96b}
found a clear correlation between the diffuse H$\alpha$ intensity
and the surface brightness of \ion{H}{2} regions on both large and
small scales. This correlation was regarded to be evidence that the
Lyc leakage is the dominant source responsible for producing the diffuse
H$\alpha$ photons. However, as we demonstrated, the radiative transfer
models for the dust-scattered halos around single \ion{H}{2} regions
can reproduce the basic properties of the H$\alpha$ halos surrounding
\ion{H}{2} regions in face-on galaxies. An advantage of the dust-scattering
scenario investigated in the present study over the in situ ionization
scenario would be the fact that the diffuse H$\alpha$ fraction of
$\approx50$\% can be explained naturally.

Since the analysis was based on the halos surrounding single \ion{H}{2}
regions, it is necessary to investigate if dust scattering can indeed
explain the global H$\alpha$ morphology, which may be an overlapping
of the dust-scattered halos surrounding individual \ion{H}{2} regions,
of the face-on galaxies. In this regard, we point out that the analyses
of \citet{Zurita02} and \citet{Seon2009}, which were initially intended
to explain the global H$\alpha$ morphology in terms of photoionization
models, can be interpreted as accounting for the morphology by the
overlapped dust-scattered halos. \citet{Zurita02} attempted to model
the diffuse H$\alpha$ morphology seen in face-on galaxies assuming
that the Lyc leakage from bright \ion{H}{2} regions would be the
predominant source of the diffuse H$\alpha$. They assumed that a
constant fraction of the Lyc leaks from bright \ion{H}{2} regions
and transferred through the two dimensional ISM disk with no significant
absorption, and claimed that the diffuse H$\alpha$ surface brightness
of NGC 157 can be reproduced reasonably well with this scenario. \citet{Seon2009}
extended the analysis to three dimensions assuming an exponential
disk, and performed a similar analysis for a face-on galaxy, M 51.
We then concluded that the effective hydrogen density required to
explain the diffuse H$\alpha$ emission with the ``standard'' scenario
was too small ($\sim10^{-5}$ of the generally known value) to be
reconciled with the effective density inferred through the turbulence
properties of ISM. In other words, unlikely extreme topology of the
ISM is required to explain the diffuse H$\alpha$ morphology by photoionization
in a turbulent or clumpy medium.

These analyses, however, did not entail self-consistent photoionization
modeling, as noted in \citet{Wood2010}. The photoionization model
should be iteratively performed until the ionization fraction and
temperature at each points converge. Instead, the penetration of photons
through ISM was characterized using an $\exp(-\tau)$ factor, after
being diluted by a geometric factor $1/r^{2}$, where $\tau$ is the
effective optical depth, and the photons interacting at a point were
assumed to be re-emitted isotropically. This radiative transfer process
in fact describes a single dust-scattering with an arbitrary dust
extinction cross-section $\sigma_{{\rm eff}}$ and an isotropic scattering
phase function (asymmetry phase factor $g=0$). In our simulations,
it was found that the H$\alpha$ luminosity from the diffuse region
(DIG) is approximately half of the total H$\alpha$ luminosity originating
from \ion{H}{2} regions. In the context of a single dust-scattering,
this is equivalent to assuming the dust albedo of $a\sim0.5$. From
the analysis, it was also found that the effective cross-section per
hydrogen nucleus is given by $\sigma_{{\rm eff}}\approx(2-4)\times10^{-22}$
cm$^{2}$ for a reasonable hydrogen density. Note that the obtained
effective cross-section is a factor of $\approx10^{-5}$ smaller than
the photoionization cross-section of Lyc, but accords well with the
dust-extinction cross-section of H$\alpha$ photons of $\sigma_{{\rm ext}}=3.8\times10^{-22}$
cm$^{2}$/H \citep{Draine03}. The absorption coefficient obtained
by \citet{Zurita02} is also consistent with the value. Therefore,
the results indicate that the diffuse H$\alpha$ morphologies in the
face-on galaxies accord well with the dust-scattered halos of H$\alpha$
photons, contrary to the claim of \citet{Zurita02} that the analysis
supports the photoionization by Lyc leaked out of bright \ion{H}{2}
regions.

In Section 2, we found that the surface brightness of the individual
dust-scattered halos can be approximated well by the formula $I(r)\propto\exp(-\tau)/r^{2}$,
which is in fact the very expression used in \citet{Zurita02} and
\citet{Seon2009}. These results seem to strongly suggest the dust-scattering
origin of the diffuse H$\alpha$ emission in the face-on galaxies.
Obviously, the radial profile is not consistent with those of the
radiation-bounded \ion{H}{2} regions. However, it is not clear at
this stage whether or not the radial profile of the photoionized H$\alpha$
emission line that is predicted from the density-bounded models accords
with $I(r)\propto\exp(-\tau)/r^{2}$. We plan to examine the radial
profiles of density-bounded \ion{H}{2} regions in the future. We
also need to develop more realistic models of the global H$\alpha$
morphologies of face-on galaxies in the contexts of not only dust
scattering, but also photoionization.

\subsection{Optical Line Ratios}

The {[}\ion{N}{2}{]}/H$\alpha$ and {[}\ion{S}{2}{]}/H$\alpha$
line ratios are observed to be higher in the diffuse H$\alpha$ regions
than in bright \ion{H}{2} regions \citep{Haffner2009}. This was
believed to be the strongest evidence against the dust-scattering
origin of the diffuse H$\alpha$ emission, because the dust scattering
would keep the line ratios constant. The enhanced optical line ratios
outside of \ion{H}{2} regions have been generally explained by a
dilute radiation field transferred from O stars within \ion{H}{2}
regions to the diffuse ISM \citep{Mathis1986,Sokolowski91,Domgorgen94}.

However, we note that the argument to reject dust scattering has been
based solely on comparisons with the line ratios from bright \ion{H}{2}
regions. If the line ratios from the ionized nebulae due to the late
O- or early B-type stars are compared with the typical line ratios
in the diffuse H$\alpha$ regions, the present scenario seems to explain
the higher line ratios than those in bright \ion{H}{2} regions equally
as well as the previous photoionization models do. It was also found
that the {[}\ion{S}{2}{]}/H$\alpha$ ratios observed in \ion{H}{2}
regions around late-type stars tend to be higher than those found
around early-type stars (Figure 3 of \citealp{Reynolds88}). \citet{ODell2011}
pointed out a striking similarity of the line ratios between Barnard\textquoteright{}s
Loop, the Orion\textendash{}Eridanus Bubble, and the typical WIM samples,
which can be explained by the photoionized nebulae due to stars with
temperature of $\lesssim35,000$ K corresponding to late O-types (see
also, \citealt{Seon2011b}). This also supports the present conclusion.

Weak {[}\ion{O}{3}{]} $\lambda$5007/H$\alpha$ and \ion{He}{1}/H$\alpha$
emission line ratios in the galactic disks also indicate that the
spectrum of the diffuse interstellar radiation field is significantly
softer than that from the average Galactic O star population \citep{Reynolds95,Madsen06}.
These line ratios provide the strongest constraint on the spectral
type of the original ionizing source of the diffuse H$\alpha$ and
other optical emission lines. However, {[}\ion{O}{3}{]} $\lambda$5007/H$\beta$
or {[}\ion{O}{3}{]} $\lambda$5007/H$\alpha$ in several halos of
edge-on galaxies are found to increase with height above the galactic
plane \citep{TuellmannDettmar2000,MillerVeilleux2003}. The rise may
indicate that other mechanisms are needed.

\citet{Seon2011b} estimated how a large portion of the ionizing power
required for the H$\alpha$ background in our Galaxy can be provided
by late OB stars, and concluded that O9 and later-type stars can account
for at least one-half of the required recombination rate of $r_{{\rm G}}\approx2.5\times10^{6}$
s$^{-1}$ cm$^{-2}$. We note that the earliest stellar type consistent
with the observed \ion{He}{1}/H$\alpha$ line ratio is O8 from our
analyses, and the fraction of Lyc luminosities of O8 and O9 stars
in the total Lyc luminosity due to O stars is $\sim18$\% (Table 1
in \citealp{Terzian1974}). Including O8 stars, late OB stars can
provide the recombination rate of $(2.3-6.2)\times10^{6}$ s$^{-1}$
cm$^{-2}$. It is therefore clear that most of the required ionizing
power in the Galaxy would be explained with late OB stars. In fact,
there is a relative lack of early O stars in the solar neighborhood,
as noted in \citet{Brandt2012}. Therefore, the diffuse optical emission
lines in our Galaxy are consistent with those produced by late OB
stars and their scattered light from the local ISM.

In our models using the MOCASSIN code, the typical values of the {[}\ion{N}{2}{]}/H$\alpha$
and {[}\ion{S}{2}{]}/H$\alpha$ ratios were well reproduced with
most of the abundances except the ISM abundances. With the CLOUDY
code, abundances close to those of WNM were required. \citet{Sembach2000}
and \citet{Mathis2000} also reproduced the observed line ratios assuming
the WNM abundances. In our previous paper \citep{Seon2011b}, we mistakenly
stated that the elemental abundances required to explain the line
ratios in the WIM are close to the B star abundances both in \citet{Sembach2000}
and in \citet{ODell2011}. This should be read as the WNM abundances
instead of the B star abundances.

The NLTE stellar models may produce significantly different amount
of Lyc photons from the LTE models. Nevertheless, we obtained consistent
results from two different model atmospheres (Kurucz or WMBASIC) in
Section 3.2. Note that it is the combination of the line blanketing
and spherical geometry, not exclusively the NTLE aspect, that is crucial
in determining the spectral energy distribution (SED) and Lyc luminosity
of stellar models \citep{Aufdenberg1998}. \citet{Aufdenberg1999}
also found that a significant difference between the spherical and
plane-parallel models is found only at surface gravities below $\log g_{*}=3.5$,
corresponding to giant stars. We also note that the calibration scale
of stellar parameters (temperature and gravity) as a function of spectral
type varies when different model atmosphere codes are used \citep{Martins2005,Simon-Diaz2008}.
\citet{Simon-Diaz2008} investigated the impact of the modern model
atmospheres (including plane-parallel stellar models) of dwarf stars
on \ion{H}{2} regions. They concluded that the predicted SEDs in
the energy range of $13-30$ eV, which are most important in producing
H$\alpha$, {[}\ion{N}{2}{]}, {[}\ion{S}{2}{]}, and \ion{He}{1}
lines, are in good agreement between the codes when the models are
compared using the calibration scales appropriate to each code. Therefore,
our main results obtained by using dwarf stars are not sensitive to
the adopted atmosphere models. Since the number of giant stars is
relatively smaller than that of dwarfs, it is also unlikely that the
main conclusions will be significantly altered, even when giants are
taken into account. Fully three-dimensional simulations with a set
of NLTE input spectra, including the effect of giant stars, may be
needed for a more detailed study.

The results of \citet{Topasna1999} are worth noting in attempting
to understand the line ratios of the diffuse H$\alpha$ regions. The
{[}\ion{S}{2}{]}/H$\alpha$ ratio observed in the Rosette Nebula
increased from $\lesssim0.2$ to $\sim0.5$ with radius out to the
boundary of the \ion{H}{2} region at $r\lesssim35'$, which was identified
by polarization and 4850 MHz radio continuum emission. Beyond the
\ion{H}{2} region boundary ($r\approx35'-60'$) where dust scattering
is dominant, there were no large variations in the ratio as a dust-scattered
halo would preserve the line ratio at the \ion{H}{2} region boundary
(see Fig.~4.27 of \citealp{Topasna1999}). The $\zeta$ Oph \ion{H}{2}
region also shows the same trend in the {[}\ion{N}{2}{]}/H$\alpha$
and {[}\ion{S}{2}{]}/H$\alpha$ ratios (Fig.~2 of \citealp{Wood2005}).
This trend is exactly what is expected from the dust-scattered halo.
However, farther out beyond these regions with constant ratios, the
line ratios increased to higher values. These higher ratios may indicate
the existence of other sources close to the nebulae to elevate the
line ratios, as will be discussed in the following paragraph.

It is thought that the line ratios increase as the H$\alpha$ intensity
decreases. This trend does not appear to be consistent with the dust
scattering scenario, because dust scattering should preserve the ratios.
We therefore need to explain the trend in the context of the dust
scattering scenario. Note that early O stars in association are more
concentrated at the galactic plane. Near an early O star, the dust-scattered
H$\alpha$ halo would be dominated by the \ion{H}{2} region due to
the O star and the H$\alpha$ intensity will be very high. The {[}\ion{N}{2}{]}/H$\alpha$
and {[}\ion{S}{2}{]}/H$\alpha$ ratios would be then relatively low
in the region. At distances far from the O star, the H$\alpha$ intensity
will be dominated by the dust-scattered light originating from the
\ion{H}{2} regions surrounding late OB stars, which are more abundant
than early O stars, and the line ratios will then be increased. As
we demonstrated using the clumpy models, some part of the enhancement
or fluctuation in the line ratios may be attributed to the variation
of clumpiness from sightline to sightline. Note that the {[}\ion{S}{2}{]}/H$\alpha$
ratio of $\sim0.23$ in the $\zeta$ Oph \ion{H}{2} region \citep{Wood2005}
is close the ratio expected from the C1 model, while the {[}\ion{S}{2}{]}/H$\alpha$
ratio of $\sim0.5$ in the Rosette Nebula \citep{Topasna1999} needs
a clumpier model than the C2 case. In LDN\,1780, the ratio of $\sim0.16$
\citep{Witt2010} is somewhere in the range between the U and C1 model.
However, the highest line ratios may need other explanations. The
highest line ratios can result from non-ionizing heating sources,
as discussed in \citet{Seon2011b}, and/or from the underlying Balmer
absorption line in the stellar continuum of relatively late-type stars,
as discussed in Section 4.

The heating sources to elevate the line ratios may include shocks,
photoelectric heating, turbulent mixing layers, and/or galactic fountain
gas \citet{Reynolds_Cox1992,Slavin1993,Collins_Rand2001,Raymond1992,Shapiro1993}.
Recently, \citet{Flores-Fajardo2011} proposed hot low-mass evolved
stars (HOLMES), which may be plentiful in the thick disks and lower
halos of galaxies, to explain the observed increase of {[}\ion{O}{3}{]}/H$\beta$,
{[}\ion{O}{2}{]}/H$\beta$ and {[}\ion{N}{2}{]}/H$\alpha$ with
increasing distance to the galactic plane in an edge-on galaxy, NGC\,891.
We also note that the \ion{H}{2} region associated with the B0.5+sdO
binary $\phi$ Per is elevated in {[}\ion{S}{2}{]}/H$\alpha$ \citep{Madsen06}.
These stars produce a much harder radiation field than massive OB
stars, and are therefore able to produce \ion{H}{2} regions with
high electron temperatures. It is possible that all these supplementary
sources may play some roles in enhancing the line ratios and their
contributions might significantly vary from sightline to sightline.
The optical emission lines due to these sources providing the highest
line ratios should also be scattered into more extended regions than
those originally produced by the sources. It should be noted that
the conventional photoionization models by O stars in the galactic
plane also need these heating sources to explain the highest ratios
\citep{Reynolds1999}. Therefore, the present dust scattering scenario
can be an alternative explanation for the diffuse optical H$\alpha$
emission that is equally as plausible as the in situ ionization scenario.

In Section 4, we argued that the {[}\ion{N}{2}{]}/H$\alpha$ line
ratio (probably, {[}\ion{S}{2}{]}/H$\alpha$ as well) in the diffuse
ISM regions seems to be consistent with that in bright \ion{H}{2}
regions, if the underlying Balmer absorption lines are taken into
account. It was also demonstrated that the line ratios should indeed
increase at the faint H$\alpha$ regions outside bright \ion{H}{2}
regions, if the diffuse H$\alpha$ emission is predominantly caused
by dust scattering. We note that a similar argument to explain the
anti-correlation of the line ratios with H$\alpha$ intensity can
be applied even to the in situ photoionization scenario. Then, the
non-ionizing heating sources proposed to explain the highest line
ratios may not be required anymore, regardless of what would be the
main origin of the diffuse H$\alpha$ emission, dust-scattered light
or in-situ ionized gas. However, the dust scattering origin is more
preferable if this is indeed the case.

\section{Summary}

We demonstrated that the basic morphological properties of the diffuse
H$\alpha$ emission can be well explained by the dust-scattered halo
surrounding the \ion{H}{2} nebula, as opposed to the previous belief
that dust scattering does not accord with the H$\alpha$ and \emph{R}-band
observations. We also found that the optical line ratios of \ion{He}{1}/H$\alpha$,
{[}\ion{N}{2}{]}/H$\alpha$ and {[}\ion{S}{2}{]}/H$\alpha$ observed
in the diffuse ISM outside of bright \ion{H}{2} regions can be well
reproduced by the dust-scattered halo scenario, wherein the optical
lines originate from \ion{H}{2} regions ionized by late O- or early
B-type stars in the media with abundances close to WNM and are scattered
off by the interstellar dust. It was also shown that the predicted
{[}\ion{N}{2}{]}/H$\alpha$ and {[}\ion{S}{2}{]}/H$\alpha$ line
ratios increase with the clumpiness of ISM. We also showed that the
Balmer absorption lines in the underlying stellar continuum seem to
explain the rise of line ratios in the faint H$\alpha$ regions. We
therefore conclude that the dust-scattering origin of the diffuse
optical emission lines cannot be ruled out. Instead, the dust scattering
can explain the observed properties of the diffuse optical emission
lines very well.

\acknowledgements{This research has made use of the NASA/IPAC Extragalactic Database
(NED) which is operated by the Jet Propulsion Laboratory, California
Institute of Technology, under contract with the National Aeronautics
and Space Administration.}

\clearpage{}

\begin{table*}
\caption{\label{scattered}Scattered fractions for various models}

\centering{}%
\begin{tabular}{ccccccccccccccc}
\hline 
\hline  & \multicolumn{5}{c}{scattered intensity / total intensity} & \multicolumn{5}{c}{scattered flux of halo} & \multicolumn{4}{c}{scattered flux over all regions}\tabularnewline
model & \multicolumn{5}{c}{at source center} & \multicolumn{5}{c}{/ total flux over all regions} & \multicolumn{4}{c}{/ total flux over all regions}\tabularnewline
\cline{2-5} \cline{7-10} \cline{12-15} 
 & 5 cm$^{-3}$ & 10 cm$^{-3}$ & 15 cm$^{-3}$ & 20 cm$^{-3}$ &  & 5 cm$^{-3}$ & 10 cm$^{-3}$ & 15 cm$^{-3}$ & 20 cm$^{-3}$ &  & 5 cm$^{-3}$ & 10 cm$^{-3}$ & 15 cm$^{-3}$ & 20 cm$^{-3}$\tabularnewline
\hline 
sphere 10 pc & 6.93\% & 12.5\% & 18.4\% & 24.1\% &  & 41.2\% & 56.9\% & 65.5\% & 71.2\% &  & 45.5\% & 63.0\% & 72.7\% & 79.0\%\tabularnewline
sphere 20 pc & 11.5\% & 21.9\% & 31.1\% & 39.4\% &  & 37.5\% & 51.1\% & 58.3\% & 62.9\% &  & 45.4\% & 62.8\% & 72.5\% & 78.8\%\tabularnewline
sphere 40 pc & 18.9\% & 33.9\% & 45.4\% & 54.5\% &  & 31.4\% & 41.5\% & 45.8\% & 48.0\% &  & 45.1\% & 62.2\% & 71.7\% & 77.7\%\tabularnewline
shell 10 pc & 9.10\% & 17.5\% & 25.2\% & 32.2\% &  & 41.3\% & 56.9\% & 65.6\% & 71.3\% &  & 45.5\% & 63.0\% & 72.7\% & 79.0\%\tabularnewline
shell 20 pc & 18.1\% & 32.4\% & 43.7\% & 52.7\% &  & 37.8\% & 51.6\% & 58.9\% & 63.5\% &  & 45.4\% & 62.8\% & 72.4\% & 78.7\%\tabularnewline
shell 40 pc & 28.2\% & 45.8\% & 57.0\% & 64.7\% &  & 32.5\% & 43.0\% & 47.5\% & 49.7\% &  & 44.9\% & 61.9\% & 71.2\% & 77.1\%\tabularnewline
\hline 
\end{tabular}
\end{table*}

\begin{table}[t]
\caption{\label{abundance}Reference abundances}

\centering{}%
\begin{tabular}{ccccccc}
\hline 
\hline Element & WNM & WNM2 & ISM & Orion & Bstar & Solar\tabularnewline
\hline 
H & 12.00 & 12.00 & 12.00 & 12.00 & 12.00 & 12.00\tabularnewline
He & 11.00 & 11.00 & 11.00 & 11.00 & 11.00 & 11.00\tabularnewline
C & 8.15 & 8.15 & 8.40 & 8.48 & 8.27 & 8.39\tabularnewline
N & 7.88 & 7.88 & 7.90 & 7.85 & 7.80 & 7.93\tabularnewline
O & 8.50 & 8.50 & 8.50 & 8.60 & 8.56 & 8.69\tabularnewline
Ne & 8.07 & 8.07 & 8.09 & 7.78 & 8.20 & 8.00\tabularnewline
S & 7.07 & 7.27 & 7.51 & 7.00 & 6.97 & 7.27\tabularnewline
\hline 
\end{tabular}
\end{table}

\begin{table}[t]
\caption{\label{sp_type}Spectral types, temperatures $T_{*}$, radii $R_{*}$,
hydrogen ionizing luminosities $Q({\rm H}^{0})$, helium ionizing
luminosities $Q({\rm He}^{0})$, and gravities $g_{*}$ of ionizing
stars}

\centering{}%
\begin{tabular}{cccccc}
\hline 
\hline Sp. Type & $T_{*}$ (K) & $R_{*}$ (cm) & $Q$(H$^{0}$) (s$^{-1}$) & $Q$(He$^{0}$) (s$^{-1}$) & $\log$ $g_{*}$\tabularnewline
\hline 
O3V & 48,410 & 1.24$\times10^{12}$ & 9.73$\times10^{49}$ & 2.35$\times10^{49}$ & 3.98\tabularnewline
O4V & 45,180 & 1.13$\times10^{12}$ & 5.46$\times10^{49}$ & 1.14$\times10^{49}$ & 3.94\tabularnewline
O5V & 42,160 & 1.03$\times10^{12}$ & 2.93$\times10^{49}$ & 4.95$\times10^{48}$ & 3.90\tabularnewline
O6V & 39,350 & 9.39$\times10^{11}$ & 1.62$\times10^{49}$ & 2.04$\times10^{48}$ & 3.86\tabularnewline
O7V & 37,150 & 8.37$\times10^{11}$ & 8.63$\times10^{48}$ & 7.32$\times10^{47}$ & 3.85\tabularnewline
O8V & 35,480 & 7.29$\times10^{11}$ & 4.30$\times10^{48}$ & 1.92$\times10^{47}$ & 3.87\tabularnewline
O9V & 33,490 & 5.92$\times10^{11}$ & 1.48$\times10^{48}$ & 2.16$\times10^{46}$ & 3.95\tabularnewline
B0V & 31,620 & 5.04$\times10^{11}$ & 4.69$\times10^{47}$ & 2.52$\times10^{45}$ & 4.00\tabularnewline
B1V & 26,600 & 4.10$\times10^{11}$ & 1.77$\times10^{46}$ & 6.70$\times10^{42}$ & 4.00\tabularnewline
\hline 
\end{tabular}
\end{table}

\clearpage{}

\end{document}